\documentclass[twocolumn]{emulateapj}
\pdfoutput=1
\usepackage{graphicx}
\usepackage{natbib}
\usepackage{multirow}
\usepackage{textcomp}

\newcommand{\bc}{\begin{center}}
\newcommand{\ec}{\end{center}}

\newcommand{\kms}{km~s$^{-1}$}
\newcommand{\cm}{cm$^{-3}$}
\newcommand{\msun}{M$_{\odot}$}

\newcommand{\am}{NH$_{3}$}
\newcommand{\htwo}{H$_{2}$}

\begin{document}

\slugcomment{}

\title{Hot Molecular Gas in the Circumnuclear Disk}

\author{Elisabeth A.C. Mills}
\affil{Department of Astronomy, Boston University, 725 Commonwealth Ave, Boston, MA 02215, USA}
\affil{Department of Physics and Astronomy, San Jose State University, 1 Washington Square, San Jose, CA 95192, USA}
\email{eacmills@bu.edu}

\author{Aditya Togi}
\affil{Department of Physics and Astronomy, The University of Texas at San Antonio, One UTSA Circle, San Antonio, TX 78249, USA }
\affil{Ritter Astrophysical Research Center, University of Toledo,
2825 West Bancroft Street, M. S. 113, Toledo, OH 43606, USA}

\author{Michael Kaufman}
\affil{Department of Physics and Astronomy, San Jose State University, 1 Washington Square, San Jose, CA 95192, USA}


\begin{abstract}
We present an analysis of archival ISO observations of \htwo\, for three 14$''\times20''$ pointings in the central 3 parsecs of the Galaxy: toward the Southwest region and Northeast region of the Galactic center Circumnuclear Disk, and toward the supermassive black hole Sgr A*. We detect pure rotational lines from 0-0 S(0) to S(13), as well as a number of rovibrationally excited transitions. Using the pure rotational lines, we perform both fits to a discrete temperature distribution (measuring up to three temperature components with T= 500-600 K, T= 1250-1350 K, and T $>$2600 K) and fits to a continuous temperature distribution, assuming a power-law distribution of temperatures. We measure power law indices of n = 3.22 for the Northeast region and n = 2.83 for the Southwest region. These indices are lower than those measured for other galaxies or other Galactic center clouds, indicating a larger fraction of gas at high temperatures. We also test whether extrapolating this temperature distribution can yields a reasonable estimate of the total molecular mass, as has been recently done for H$_2$ observations in other galaxies. Extrapolating to a cutoff temperature of 50 K in the Southwest (Northeast) region, we would measure 32 (140) \% of the total molecular gas mass inferred from the dust emission, and 26 (125) \% of the total molecular gas mass inferred from the CO emission. Ultimately, the inconsistency of the masses inferred in this way suggest that a simple application of this method cannot yield a reliable estimate of the mass of the Circumnuclear Disk.
\end{abstract} 


\section{Introduction}
The center of our Galaxy hosts a supermassive black hole with a mass of 4$\times10^6$ M$_{\odot}$ at a distance of $\sim$ 8 kpc \citep{Boehle16}, which is currently accreting in an extremely quiescent state, having a bolometric luminosity many orders of magnitude below the Eddington luminosity at this mass \citep{Narayan98,Baganoff03}. Surrounding both the black hole Sgr A* and a central nuclear star cluster \citep{Lu13}, with an inner radius of 1-1.5 pc, is a molecular gas torus known as the circumnuclear disk \citep[CND, e.g.,][]{Genzel85,Gusten87}. This is the closest reservoir of molecular gas to the black hole, and represents the gas available for its future feeding, activity, and associated star formation. Estimates of its mass vary, but recent studies have converged on a value of a few $10^4$ M$_{\odot}$ \citep{Etx11,RT12}. The properties of this gas are much more extreme than those seen in typical molecular clouds: it has broad line widths ($\sigma \sim$ 10-40 \kms\,) on 5$''$ (0.2 pc) scales indicative of strong turbulence \citep{Chris05,MCHH09}, as well as high average densities of $10^5-10^6$ \cm \citep{RT12,Mills13}. Some of the molecular gas is also being ionized by the nuclear star cluster \citep{Zhao93}. Additionally, compared to the cosmic ray ionization rate in the solar neighborhood \citep[ $\zeta\sim3\times10^{-16}$,][]{Indriolo12} $\zeta$ is likely at least three times higher \citep[ estimates range from $>10^{-15}$ to $2\times10^{-14}$][]{Goto08,Goto13,Harada15}. All together, it is perhaps the most extreme environment in our Galaxy for forming stars. 

Whether or not the CND is currently forming stars is controversial. Measurements of density indicate that the broad linewidths measured by \cite{Chris05} and \cite{MCHH09} are highly supervirial, and the clumps are therefore not self-gravitating \citep{Smith14}. Despite this, there have been suggestions of star formation based on several indicators, including shock-excited methanol masers and candidate outflows traced by SiO (5-4) \citep{YZ15b}, as well as compact, highly-excited SiO emission interior to the CND \citep{YZ13d}. However, the bulk of these have other plausible explanations: methanol masers can be excited in strong shocks and are ubiquitous throughout the Galactic center in the absence of other signs of star formation \citep{YZ13, Mills15}, and in fact many of the masers highlighted by \cite{YZ15b}, are at velocities not associated with CND gas. Linewidths in the CND, as noted above, are generally broad with complex profiles that can be attributed to extreme turbulence in this source. Finally, in the vicinity of the strong radiation field from a nuclear star cluster, the detection of highly-excited SiO may indicate radiative excitation \citep{Godard13} rather than denser gas that could be associated with a protostar. At present then, conclusive evidence for active star formation in the CND is lacking, and this structure appears, like many other Galactic center clouds \citep[e.g.,][]{Longmore13} to be relatively quiescent.

Although observations motivated by constraining star formation have led to estimates of the gas density, we are still lacking the full picture of the physical conditions that dominate in this region. In particular, new observations have failed to yield improved estimates of the gas temperature in this region. The temperature is unconstrained in the analysis of \cite{Mills13} using HCN and HCO$^+$ over a range from 50 to 300 K. While \cite{Bradford05} report a temperature of $\sim$200-300 K using CO, they acknowledge that there are no stringent upper limits on this value. This is consistent with earlier observations of \cite{Lugten87}, as well as observations of a larger number of CO transitions by \cite{RT12} that constrain the temperature only to be $\gtrsim$150 K, over a range of considered temperatures that extends up to 600 K. While \cite{HH02} detect the highly-excited \am\, (6,6) line (E$_{upper}\sim$ 400 K), they do not calculate a gas temperature from their observations. In comparison, many more observational constraints on the dust temperature exist: dust temperatures as warm as 90 K have been measured, though these are only suggested to apply to a small fraction ($<10\%$) of the CND mass \citep{Etx11, Lau13}. However, this may not give any constraint on the dust temperature, as the gas and dust temperatures are observed to be decoupled in Galactic center clouds \citep[e.g.,][]{Ginsburg16}, and the gas densities present may not be high enough that the gas and dust would be expected to be thermalized here, given the high cosmic ray ionization rate \citep{Clark13, Goto13}. 

Inside the central cavity of the CND, the gas is predominately atomic \citep{Jackson93}, with an estimated mass of 300 M$_{\odot}$, and its temperature is also not constrained However, there is also an ionized component \citep[the `minispiral'][]{Lo83,Ekers83} with about 10\% of the mass of the atomic component, for which the temperature of the intermixed dust is measured to be as high as 220 K \citep{Cotera99b}. There is also suggested to be a hot molecular component, as highly-excited CO is detected toward Sgr A* with Herschel \citep{Goic13} that is consistent with temperatures of $\sim$1300 K. Resolved observations of the inner edge and cavity of the CND in rovibrational lines of \htwo\, have been made by \cite{Ciurlo16}, and indicate temperatures up to a few thousand K, with the total mass of gas at these temperature being $<10^{-2}$ M$_{\odot}$. However, much of the cavity gas also shows strong deviations from thermal equilibrium. 

Accurate gas temperature measurements in the central parsecs are critical for advancing our understanding of the CND on several fronts. First, improved constraints on the temperature are needed for better constraints on the density. Without strong constraints on the gas temperature, the degeneracy between high temperature/low density and low temperature/high density solutions to radiative transfer models leads to orders of magnitude uncertainty in density measurements \citep[e.g,][]{RT12,Mills13,Smith14}. More precise measurements of the gas density structure are needed in order to better assess the evolution, longevity, and star-forming potential of this structure. Improved constraints on the temperature are also needed to determine the dominant heating mechanism for the dense molecular gas in the CND. Cosmic rays \citep{Goto08,Harada15}, UV radiation \citep{Lau13,Ciurlo16}, X-rays \citep{Goto13}, and turbulent dissipation \citep{Lugten87} have all been suggested to contribute to the heating in this environment. Isolating the dominant heating source is relevant for determining the extent to which the CND gas may be taken to be an analog for gas in the centers of more extreme galaxies like ULIRGs, nuclear starbursts, or even AGN. 

In clouds in the central 300 pc of the Galactic center, most gas temperature measurements have been made with \am\,\citep{Gusten85,Huttem93b,MM13}, other symmetric tops like CH$_3$CN and CH$_3$CCH \citep{Gusten85}, CO \citep{Martin04}, and more recently H$_2$CO \citep{Ao13,Ginsburg16}. However, in the CND these tracers have not yielded well-defined temperatures. H$_2$CO is extremely weak in the CND, likely due to photodissociation \citep{Martin12}, and the same is true of CH$_3$CN and CH$_3$CCH (Riquelme, private communication). As already noted, analysis of CO and \am\, have also failed to yield well-defined constraints on temperature \citep{Lugten87, RT12, HH02}. In this paper, we undertake an alternative approach by directly measuring the temperature of the gas using \htwo. Typically, due to a combination of the lack of a permanent dipole moment of \htwo\, and the relatively high energy of the lowest rotational transition (510 K), \htwo\, is not a detectable tracer of the cool molecular ISM, and one must rely on the previously-listed indirect tracers of gas temperature. However, clouds in the Galactic center are much warmer, and so these lines can be detected in a significant fraction of the gas. As an example, \cite{RF01} detect pure-rotational lines of \htwo\ in a sample of 16 Galactic center clouds, and measure gas with temperatures from 150-600 K that comprises 30\% of the total \htwo\, column. 

In this paper, we undertake an analysis of archival ISO spectra of pure-rotational and rovibrational lines of \htwo\, in three positions toward the CND and central parsec. There are a number of prior observations of rovibrational lines in the near-infrared \citep{YZ01, Lee08, Feldmeier14,Ciurlo16}, however to our knowledge this is the first analysis of the pure-rotational lines in this source. We present a temperature analysis of these line that includes both a fit to a discrete number of temperature components, as well as a power-law analysis assuming a continuous distribution of temperatures. We then discuss the fraction of warm \htwo\, that is detected in the CND, the implications of this temperature distribution for identifying a heating source, and the uniqueness of the CND compared to other sources in which \htwo\, temperatures have been measured.

\section{Data}
\label{dat}
The data used in the analysis in this paper were obtained from the NASA/IPAC Infrared Science Archive. We analyzed spectra from three positions observed with the Infrared Space Observatory \citep[ISO;][]{Kessler96} toward the central two parsecs of the Milky Way. All observations were made in May 1996 using the Short Wavelength Spectrograph \citep[SWS;][]{deGraauw96} in low-resolution, full-grating scan mode. Spectra of the central pointing toward the black hole Sgr A* (TDT number 09401801, RA = 17$^{\mathrm{h}}$45$^{\mathrm{m}}$39.97$^{\mathrm{s}}$, Dec= $-29\degr00'28.7''$) were published in \cite{Lutz}, however the H$_2$ lines were not analyzed. The other two pointings, toward the Northeast region (TDT number 09401504, RA=17$^{\mathrm{h}}$45$^{\mathrm{m}}$41.76$^{\mathrm{s}}$, Dec= $-28\degr59'50.7''$) and the Southwest region (TDT number 09401905, RA= 17$^{\mathrm{h}}$45$^{\mathrm{m}}$38.58$^{\mathrm{s}}$, Dec= $-29\degr01'05.8''$) have not been published. 

\begin{figure}[tbh]
\includegraphics[scale=1]{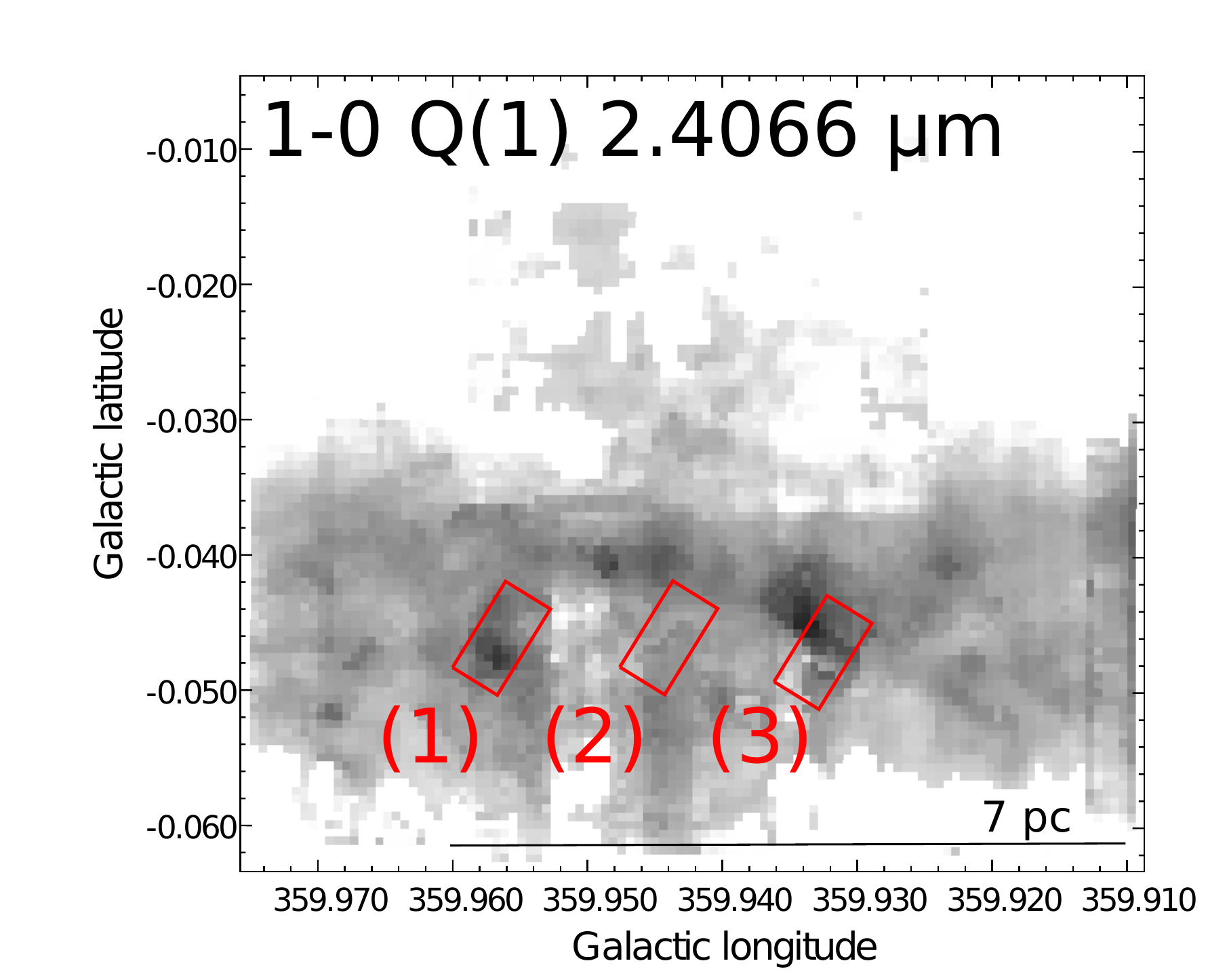}
\caption{Locations of the ISO SWS apertures for the observed \htwo\, spectra, superposed on a map of the 1-0 Q(1) line of \htwo\, taken with the near-infrared long-slit spectrograph ISAAC on the VLT using drift scans \citep{Feldmeier14}. The ISAAC data have a resolution (pixel size) of 2.22$''$.The ISO SWS aperture size shown here is $14''\times20''$. Position (1) corresponds to the Northeast region of the CND, position (2) corresponds to Sgr A*, and position (3) corresponds to the Southwest region of the CND.}
\vspace{-0.7cm}	
\label{fig1}
\end{figure}

The spectra were obtained with an aperture of $14''\times20''$ for wavelengths of 2.38-12.0 $\mu$m, $14''\times27''$ for wavelengths of 12.0-27.5 $\mu$m, and $20''\times27''$ for wavelengths of 27.5-29.0 $\mu$m. Only the 0-0 S(0) H$_2$ line at a wavelength of 28.221 $\mu$m is observed with an aperture of $20''\times27''$. The 0-0 S(1) and S(2) lines are observed with an aperture of $14''\times27''$ and the remainder of the lines are observed with a consistent aperture of $14''\times20''$. However, the SWS spectra we use are highly-processed data products that have had their continuum level normalized to a consistent value \citep{Sloan03}. Based on mapped observations of the 1-0 Q(1) line \citep{Feldmeier14}, we assume that the \htwo\, emission can be taken to fill the aperture with a filling fraction of 1 for this range of aperture sizes, and so in subsequent calculations we take the effective aperture size to be $14''\times20''$, and apply no additional correction for variations in the aperture size between lines. For all aperture sizes, the long axis of the slit is oriented with a position angle of -1$\degr$.4 in an ecliptic coordinate frame. The position of these pointings is shown in Figure \ref{fig1}, superposed on the map of the 1-0 Q(1) line from \cite{Feldmeier14}

\begin{figure}[tbh]
\includegraphics[scale=0.5]{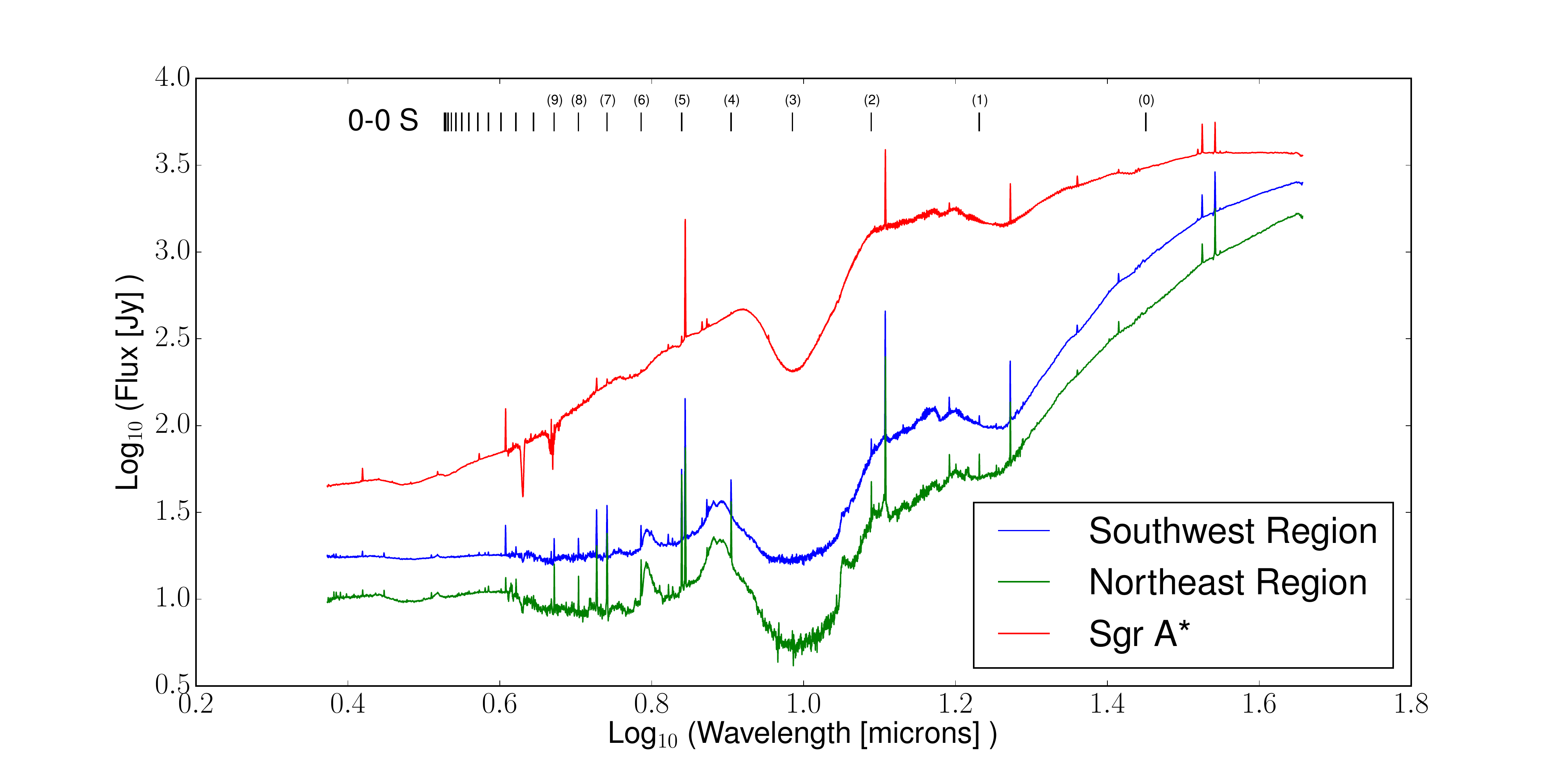}
\caption{SWS spectra for the three positions observed with ISO toward the central parsecs of the Galaxy: Sgr A*, the Northeast region of the CND, and the Southwest region of the CND. Wavelengths corresponding to the 0-0 S pure rotational lines are indicated. The aperture size changes at 12 $\mu$m, 27.5 $\mu$m and 29 $\mu$m, however the spectra have been scaled so that the continuum level is consistent across these changes. }
\vspace{-0.7cm}	
\label{fig2}
\end{figure}

The full spectra toward each position from 2.5 to 40 $\mu$m are shown in Figure \ref{fig2}. Integration times for these observations were 6528 s, resulting in per-channel noise levels of $\sim$ 0.5-1 Jy for all lines except the 0-0 S(0) line, which has a noise of $\sim$ 10 Jy. Additional baseline fluctuations with amplitudes 0.5-1 Jy are present in the spectra that can become as high as 5-40 Jy toward Sgr A*. The channel width of the observations is $\sim$ 15 \kms, however the actual instrumental spectral resolution is R$\sim$1000-3000, or 100-300 \kms. We fit a first-order polynomial to the continuum in the $\pm$2000 \kms\, surrounding the line. The continuum levels range from 10 to 3000 Jy and the shapes are generally well-fit by this approach. For the S(0) line, the continuum is larger, with a steeper slope and greater variation. For this line, we fit a first-order polynomial to a more limited range. The baseline-subtracted spectra of the 0-0 S, 1-0 Q, and 1-0 O lines are shown in Figure \ref{fig3a}, \ref{fig4}, and \ref{fig5}.

\section{Results and Analysis}
We detect multiple pure-rotational and rovibrational lines of \htwo\, toward all three positions. The strongest \htwo\, emission is detected from the North, with slightly weaker emission in the Southwest and toward Sgr A*. Toward the Southwest region we detect 12 pure-rotational lines, up to 0-0 S(13), and 7 rovibrational lines, up to 1-0 Q(5) and 1-0 O(6). Toward the Northeast region we detect 13 pure-rotational lines up to 0-0 S(13), and 10 rovibrational lines up to 1-0 Q(7) and 1-0 O(6). This is half the number of H$_2$ pure rotational and rovibrational lines that have been detected by SWS toward the Orion-KL outflow \citep[56;][]{Rosenthal00}. Toward Sgr A* we detect only 5 pure-rotational lines from S(4) to S(8), and we only detect two rovibrational lines: O(4) and O(5). The continuum is significantly stronger toward Sgr A*, and the baseline structures are more complicated, hindering the detection of a similarly large number of lines in this source. Typical baseline uncertainties are 0.3 Jy for N and S, and 0.7 Jy for Sgr A*.

We fit Gaussian profiles to all of the detected lines (shown in Figures \ref{fig3a}, \ref{fig4}, and \ref{fig5}), and report the line parameters for each source in Tables \ref{S00}, \ref{Q10}, and \ref{O10}. Typical measured linewidths are $\sim$140$\pm$40 \kms\, so considering that the spectral resolution of these SWS observations ranges from 100-300 \kms \citep{Valentijn96}, the lines are likely not resolved, and their shape is dominated by the instrumental profile. Additionally, the central velocities show variation within each source, on the order of 30-40 \kms, which can be attributed to the uncertainty in the wavelength calibration of these observations, which has been measured to range from 25-60 \kms \citep{Valentijn96}. We measure a mean central velocity of -34$\pm$38 \kms\, toward the Southwest region, 57$\pm$38 \kms\, toward the Northeast region, and 28$\pm$25 \kms\, toward Sgr A*. 

There are several lines for which it appears there is a contaminating line at nearly the same wavelength. The 1-0 O(2) line at 2.62688 $\mu$m shows two signatures indicative of contamination with a species tracing the ionized gas: it is anomalously strong compared to other nearby H$_2$ lines, and the emission toward Sgr A* is stronger than the emission toward the two pointings in the CND, which is not observed in other H$_2$ lines. In addition, the 1-0 O(2) line appears offset from the velocity inferred from the H$_2$ lines. Based on the strength and velocity of the feature near the 1-0 O(2) line, we identify this as the Brackett $\beta$ line at 2.62587 $\mu$m at an offset velocity of 114 \kms. Contamination with this line is also noted in SWS observations of NGC 1068 by \cite{Lutz00}. In addition, the wavelengths of the 1-0 Q(7) line at  2.5001$\mu$m and the 3-2 S(0) line at 2.5014$\mu$m are nearly overlapping (having an offset velocity of 183 \kms). If we assume the feature detected at this wavelength is 3-2 S(0), then the measured velocity is significantly offset from that measured for the other H$_2$ lines. We thus infer that the emission here is dominated by the 1-0 Q(7) line, and do not report a detection of the 3-2 S(0) line. There are also several lines for which there are features at nearby wavelengths that appear as emission at offset velocities (e.g., in the 1-0 O(8) and 1-0 Q(2) lines, and the 2-1 O(6) line) that are likely due to other species.

From the Gaussian fits to the flux of each line, we can calculate a column density for each level using the relation

\begin{equation}
N(v,J) = \frac{4 \pi \lambda}{h c} \frac{I_{obs}(v,J - v', J')}{A_{ul}(v,J - v', J')} e^{A(\lambda)/1.086}
\label{eq1}
\end{equation}

where $I_{obs}(v,J - v', J')$ is the observed line flux from the transition from level $(v,J )$ to $( v', J')$, $A_{ul}(v,J - v', J')$ is the Einstein A radiative transition probability from level $(v,J )$ to $( v', J')$, and $A(\lambda)$ is the extinction at the wavelength of that transition. As previously stated, the column densities calculated in this way assume a beam filling fraction of 1 for the H$_2$ emission; if the emission is clumpy, the column densities would then be a lower limit on the true value.

\subsection{Extinction Correction}
We find that the 0-0 S(3) line is anomalously weak, lying below the other 0-0 S lines in a Boltzmann plot (Figure \ref{fig6}), a result that can be attributed to extinction from the  9.7 $\mu$m silicate feature \citep[e.g.,][]{RF01,Lutz}. In order to correct the observed \htwo\, lines for the extinction at mid-infrared wavelengths we adopt the \cite{Fritz11} extinction law derived from the ISO continuum toward Sgr A*. As extinction is a complicated function of the wavelength, we describe this law by taking the values of 100 data points on their derived curve from 0.43 $\mu$m to 25$\mu$m and interpolating to obtain the extinction values at the observed wavelengths. The extinction values from \cite{Fritz11} are then scaled in order to minimize the scatter in a linear fit to the S(1), S(2), S(3) and S(4) lines. We note that the is a significant improvement over the \cite{Lutz} law adopted by \cite{RF01}. As can be seen in Figure \ref{fig6}, interpolating a fit using the same method is unable to bring the S(3) line into alignment with the S(2) and S(4) lines without assuming a much larger extinction as well as overcorrecting the S(2) line to be brighter than the S(1). We note however that the extinction correction does not significantly change the measured temperatures. 

The applied extinction correction using the \cite{Fritz11} law significantly raises the flux of the S(3) line, making it consistent with the flux observed in the other 0-0 S pure rotational lines. It indicates that the extinction in N and S is only slightly larger than that derived by \cite{Fritz11} toward Sgr A* (Note that as the extinction law is determined using the same ISO observations of Sgr A* that are analyzed here, we adopt the unscaled \cite{Fritz11} extinction values for this latter source). However, minimizing the scatter in a linear fit to the first four detected rotational lines of \htwo\, does not account for the slight curvature seen in the Boltzmann plots of the \htwo\, lines. This method thus actually slightly overestimates the extinction, yielding a slightly overlarge flux for the S(3) line. We thus find it is better to perform a simultaneous optimization of the fit to extinction and temperature, as described in the following section.

\subsection{Discrete Temperature Fitting}
We next plot the column densities $N(v,J)$ divided by the level degeneracy $g_J$ against the upper level energy $E(v,J)/k$ in a `Boltzmann plot'. For \htwo\, the level degeneracy is $g_J = g_s (2J+1)$ where $g_s$ = 3 for ortho-\htwo\, (odd $J$) and $g_s$=1 for para-\htwo\, (odd $J$). With the exception of the pointing toward Sgr A* (for which only 5 pure-rotational lines are detected), the pure-rotational lines measured toward the CND follow a convex or `positive' curve that is generally interpreted as the presence of multiple temperature components. While \cite{Neufeld12} has found that a sufficiently low-density (log $n\, <$ 4.8) isothermal gas can also reproduce a positive curvature in a Boltzmann plot for {\bf CO pure rotational line emission}, no work has yet searched for a similar effect with the H$_2$ molecule. Lacking this detailed modeling, we note that there are several reasons to suspect that this specialized case is not applicable to our H$_2$ measurements. First, the curvature induced in this way is slight, and is not sufficient to explain many of the observations examined by \cite{Neufeld12} without still appealing to multiple temperature components. Second, lower critical densities of the pure-rotational H$_2$ lines \citep{leBourlot99} suggest that substantially lower densities would be required, should this effect manifest in H$_2$ observations, while in fact the bulk of the CND gas has higher densities: recent measurements of the density of CND gas range from log $n = 4.5-6.5$ \citep{RT12,Mills13}. We thus assume for this work that the substantial curvature in the Boltzmann plots indicates that the CND gas is not isothermal.

In order to constrain multiple temperatures present in the CND gas, we first follow the approach of \cite{Rosenthal00} and perform a simultaneous fit to three temperature components in the pure-rotational lines observed in the Southwest and Northeast regions (note that with the smaller number of lines detected toward Sgr A*, we can only justify fitting a single temperature component in this source). The measured column densities are fit to a {\bf summation of three column densities from} three single-temperature components (each of which would be a straight line in the Boltzmann plot):

\begin{equation}
N(v,J) / g_j) = \sum_{i=1}^{3} C_i\, e^{-E(v,J) / kT_{ex,i}}
\label{eq1b}
\end{equation}

{\bf As before, $E(v,J)$ is the energy of level $(v,J )$, $g_j$ is the level degeneracy, and $N(v,J)$ is the column density determined using Equation \ref{eq1}. The $C_i$ are constants determined by our fitting, representing the contribution of each of the three components to the total column density, and the $T_{ex,i}$ are the temperatures of each component.} We perform a minimization of the residuals of a {\bf least-squares} fit to simultaneously determine the best-fit temperature (the inverse of the slope of the line) for each of the three temperature components {\bf as well as its contribution to} the total column density (the y-intercepts {\bf or $C_i$ for each component}). {\bf In addition to letting the $T_{ex,i}$ and $C_i$ for each component be free parameters in our fitting, we also allow the extinction value (used to calculate $N(v,J)$ from the observed line fluxes) to be a free parameter, in order to determine a more accurate extinction correction, given the curvature in the Boltzmann plot.} 

The results of this minimization are shown in Figure \ref{fig9}, {\bf and the best-fit $T_{ex,i}$, $C_i$, and $N_{\mathrm{H2},i}$ for each component are given in Table \ref{fraction}}. The extinction, and the two lowest-temperature components (the `hot' and `hotter' gas) are well-constrained by our data. The best-fit extinction is 0.98 $\times$ the extinction toward Sgr A* (or A$_{L'}$ = 1.07) for the Southwest region, and slightly higher for the Northeast region: 1.17 $\times$ the extinction toward Sgr A* (or A$_{L'}$ = 1.28). However, the highest-temperature or `hottest' component is not strongly constrained, and is effectively a lower limit on the highest temperatures present ( $\gtrsim$ 2600 K, which is is also approximately the temperature of the detected rovibrational Q(1) and O(1) lines of \htwo\, at these positions.). However, we will use the best-fit values for the very hot component in our estimates of the total column density which follow. We find best-fit temperatures of 580 K, 1350 K, and 3630 K toward the Southwest region, and 520 K, 1260 K, and 2840 K toward the Northeast region. Toward Sgr A*, we separately fit a single temperature component and find the best-fit temperature to be 1100 K.  We find that the temperatures in the Northeast region are 100-200 K cooler than in the Southwest region. One possible explanation is that this is because the Southwest region is closer to a source of heat provided by the central nuclear cluster. After all, only the Southwest edge of the CND appears to be ionized, while the Northeast edge does not \citep{Zhao93}. We discuss other alternatives in Section \ref{shock}.

The extremes of temperature that we find are higher than temperatures that have been measured in the CND with other tracers. The hottest measured dust temperatures are 220 K \citep{Cotera99b}, and measurements of highly-excited CO using Herschel by \cite{Goic13} find temperatures of $\sim$ 1300 K toward the central cavity (consistent with our detection of T=1100 K in this direction, though our detection of a single-temperature component in this region may be an observational bias, due to weaker lines and more variable spectral baselines in this region).  The only measurements of comparably-hot gas are of T$\sim$ 1700-2500 K gas measured by \cite{Ciurlo16} for other near-infrared rovibrational lines of \htwo\ in the inner edge and central cavity of the CND. However, there is a large uncertainty on the highest temperature they measure (T$\sim$ 2500$^{+3000}_{-900}$), and the majority of the positions they analyze show significant deviations from a thermal distribution that they attribute to recent formation of \htwo\ in the central cavity. In contrast, we find that the 0-0 S lines in the Northeast and Southwest regions of the CND appear to be consistent with a thermal distribution up to temperatures of at least 2600 K. 

The resulting Boltzmann diagrams of the extinction-corrected column density for each position are shown in Figure \ref{fig7}. Each individual temperature component is plotted in the Boltzmann diagram as a dashed line, with the sum of all three components plotted as a solid curve. In addition to the 0-0 S lines which are used in the temperature fit, we also plot the $v=1-0$ O and Q rovibrational lines. We find that they follow a roughly straight line on these plots, lying below the pure-rotational lines with similar upper level energies. The temperature of these lines is roughly consistent with that of the highest-temperature gas component measured in the 0-0 S lines ($\sim$ 3000 K). These two distinct distributions are in contrast to what is observed in the Orion-KL outflow in which the rotational and vibrational lines are thermalized and follow a single distribution in the Boltzmann diagram (\citealt{Rosenthal00}, although the high-J CO lines observed in Orion KL do show a curved distribution similar to the 0-0 S \htwo\, lines in the CND, e.g., \citealt{Goic15}). Although we do not detect the S(0) (J=2--0) line, we can also estimate a T$_{32}$ temperature from our measured upper limit on the strength of this line and our measurement of the S(1) (J=3--1) line. We constrain T$_{32}$ $>$110 K in the Southwest region and $>$130 K in the Northeast region. 

Using the temperature fits, we can extrapolate the contribution of each component to lower J and determine the total column density of each temperature component. The fraction of the  \htwo\, column in each component (`hot' / `hotter' / `hottest') is given in Table \ref{fraction}. We find that for the Northeast and Southwest regions, the vast majority of the detected \htwo\, is in the `hot' 500-600 K component: more than 95\% for both sources. Both sources have 3-4\% of the \htwo\, column in the `hotter' 1200-1300 K component, and less than 0.1\% of the \htwo\, column in the `hottest' T$\sim$ 3000 K component. While this makes both sources appear quite similar, the best-fit temperature components in the Northeast region are somewhat cooler than those in the Southwest region.  

Summing over all the components we can also estimate the total column of hot (T$>$500 K) \htwo\, in each source. For the Southwest region we measure a total column of 6.4$\times10^{20}$ cm$^{-2}$.  For the Northeast region we measure a total column of 1.2$\times10^{21}$ cm$^{-2}$. As the fitting uncertainties are small, the dominant error in these quantities is likely the absolute flux calibration of ISO-SWS, which is estimated to be 12-20\% for the wavelength range studied here \citep{Schaeidt96}. We thus find that there is significantly more hot \htwo\, in the Northeast region (around a factor of 2) than in the Southern region. This is consistent with our observations that the \htwo\, lines are stronger in the Northeast than in the Southwest.

\subsection{Continuous Temperature Fitting}
As we do not detect the S(0) line in the ISO data, our discrete temperature fits are not very sensitive to the presence of `warm' gas with T $<$ 500 K. Further, as the discrete fitting approach seeks to optimize a fit to three temperature components that are not consistent between the Northeast and Southwest regions, it is difficult to objectively compare the hot gas properties (such as the fraction of gas at each temperature) between these two sources. We thus also employ a second approach for quantifying the temperature differences between the Northeast and Southwest regions.

Here, we first assume that a fixed number of discrete temperature components is not a physically well-motivated model for this gas, and that a more realistic description would be gas in which the temperature varies smoothly or continuously from cool to hot values, as can be generated by a range of pre-shock densities, shock velocities, or shock geometries \citep{NY08,YN11,Appleton17}.  We then follow the approach of \cite{Togi16} by adopting a continuous power-law temperature distribution for the \htwo, which extrapolates the temperature distribution down to lower values than those we were able to measure. In this model, the measured column densities of H$_{2}$ molecules are modeled as the distribution from a power law function with respect to temperature, dN $\propto$ T$^{-n}$ dT, where dN is the number of molecules in the temperature range T$\rightarrow$T+dT. The model then consists of just three adjustable parameters: an upper and lower temperature cutoff (T$_{u}$ and T$_{\ell}$) and a power law index (n).

As our primary goals in adopting this model are to compare the power law indices between the two observed regions, and to make an estimate of the total amount of \htwo\, which could be present if this power law extends to lower temperatures (T$<$ 500 K), we focus only on the S(1) to the S(7) lines. We also keep the upper temperature, T$_{u}$, fixed at 2000 K, as hotter gas is a negligible contributor to the total column. The only two parameters which then vary in the model fitting are the lower cutoff temperature T$_{\ell}$ and the power law index n. We then use this model to fit the observed column densities for the Northeast and Southwest lobes (corrected for the best-fit extinction value that was determined in our discrete temperature fits). We report results for two adopted values of a lower cutoff temperature: T = 50 K, roughly consistent with the coolest molecular gas typically measured in other clouds in the Galactic center \citep[][]{Ao13,Ginsburg16}, and T = 100 K, which is roughly the coolest gas allowed by our nondetection of the S(0) line (though given the large upper level energy of this line, cooler gas may be present that would not contribute emission to this line). 

Figure \ref{fig8} shows the resulting model fits to the observed H$_{2}$ columns for the Southwest and Northeast regions. We find that a fit to a continuous temperature distribution shows that the Southwest region (n = 2.83) is systematically warmer than the Northeast region (n=3.22), where a shallower or smaller power law index indicates a greater fraction of warm/hot gas. This is consistent with the trend seen in our discrete temperature fits, that the Northeast region appears systematically cooler than the Southwest region. The power law indices measured for both regions of the CND are much shallower than typical values measured in the centers of other star-forming galaxies, indicating that the CND gas is hotter than typical gas in the nuclei of other galaxies \citep{Togi16}. We will discuss this further in Section \ref{compare}.

The best-fit power law for each region can then be extended to lower temperatures in order to estimate the total column of warm \htwo\, with temperatures greater than the assumed cutoff value (50 or 100 K). Adopting a cutoff value of 100 K for the coolest gas present, we measure a total column of warm \htwo\ of 1.2$\times10^{22}$ cm$^{-2}$ for the Northeast region, and 4.4$\times10^{21}$ cm$^{-2}$ for the Southwest region. In this case, there is $\sim$ 3.5 times more warm (T$>$ 100 K) \htwo\, present in the Northeast region than in the Southwest region, compared to twice as much hot (T$> 500$ K) \htwo\, in the Northeast, as measured from the discrete temperature fits. The warm \htwo\, column would also be $\sim$4-5 times larger than the hot column that we measured with our discrete temperature fits. If we go further and adopt a lower cutoff temperature of 50 K, the total column of \htwo\ present would be 5.6$\times10^{22}$ cm$^{-2}$ for the Northeast region, and 1.6$\times10^{22}$ cm$^{-2}$ for the Southwest region. This is again $\sim$ 3.5 times more warm (T$>$ 100 K)\htwo\, present in the Northeast region than in the Southwest region. The overall increase in \htwo\, column would also mean that if this power-law extrapolation holds, and if there is gas as cool as 50 K in the CND, 70-80\% of the \htwo\, would have temperatures in the range of 50-100 K. We discuss this further in Section \ref{interpret}.

\subsection{The warm/hot gas mass of the CND and the central cavity}

We can also translate these total column densities of H$_2$ into total masses within the observed ISO-SWS aperture for the extrapolated amount of warm (T$> $50 or 100 K) gas and the measured amount of hot (T$>$ 500 K) gas. The total \htwo\, mass $M_{H2}$ is given by:

\begin{equation}
M_{H2} = N_{H2}\, \Omega\, d^2 \mu_{\mathrm H2} \hspace{0.1cm} m_{\mathrm H}
\label{eq2}
\end{equation}

where $N_{H2} $ is the total measured column density of \htwo, $\Omega$ is the solid angle of the ISO-SWS aperture, $d$ is the distance to the Galactic center \citep[8 kpc;][]{Boehle16}, $\mu_{\mathrm H2}$ is the mean molecular weight of the gas, which we take to be 2.8, assuming abundances of 71\% H, 27\% He, and 2\% metals \citep[e.g.,][]{Kauffmann08}, and $m_{H}$ is the mass of a Hydrogen atom. Taking an aperture size of 14$''\times20''$ (0.54 by 0.78 pc on the sky, at the assumed distance of 8 kpc), the measured total column densities from our discrete temperature fits would correspond to a total mass of hot molecular gas of 6.1 M$_{\odot}$ in the Southwest region, and 11.4 M$_{\odot}$ in the Northeast region. 

Additionally, we can determine the mass of hot molecular gas in the central cavity from the observation toward Sgr A* (for just the measured 1100 K temperature component). Here, we find a total mass of $\sim$0.5 M$_{\odot}$ within the ISO-SWS aperture. In comparison, there is measured to be roughly 30 solar masses of ionized gas in the minispiral, and 300 solar masses of neutral gas in the central cavity\citep{Jackson93}. This is a larger mass of hot \htwo\, in the central cavity than measured by \cite{Ciurlo16}, who estimate a total \htwo\, mass of $\sim7\times10^{-3}$ M$_{\odot}$ if the emission they observe is representative of emission along a narrow inner edge of the CND that abuts the central cavity. However, unlike the high-resolution observations of \cite{Ciurlo16} and as we discuss in Section \ref{spatial}, it is not clear that all of the gas we detect toward Sgr A* is actually confined to the central cavity. There is then no requirement that this hot gas be associated (solely) with the inner edge of the CND or the central cavity. 

The total molecular mass of gas inferred to be present from our extrapolated continuous temperature fits is much larger than the mass of just the hot gas determined from the discrete temperature fits. For a cutoff temperature of 100 K, the total molecular gas mass in Southwest region would be 41 M$_{\odot}$, and for the Northeast region it would be 111 M$_{\odot}$. For a lower cutoff temperature of 50 K, the total molecular gas mass in the Southwest region would be 146$\pm$54 M$_{\odot}$, and for the Northeast region it would be 518$\pm$243 M$_{\odot}$. 

We can compare these masses to other independent estimates of the total molecular gas mass in these regions. CO observations \cite{RT12} inferred masses of 795 M$_{\odot}$ and 590 M$_{\odot}$ toward nearly identical positions in the Southwest and Northeast respectively, for a beam with FWHM = 22$''$.5. As these are just masses of \htwo\, (Requena Torres, private communication) we first scale these masses for the adopted mean molecular weight. As solid angle of this beam is approximately twice the ISO aperture, we also scale the masses to the ISO aperture, assuming uniform emission. The resulting masses for the Southwest (557 M$_{\odot}$.) and Northeast (413 M$_{\odot}$.) regions are also reported in Table \ref{masses}. We can also measure a molecular mass from the \htwo\, columns determined from Herschel dust measurements (Battersby et al., in prep. and private communication).  These \htwo\, column densities were derived by fitting a modified blackbody to each pointing in the Galactic center, with the dust temperature allowed to vary as a free parameter. The dust opacity as a function of frequency is determined by fitting a power-law to tabulated opacities from \cite {OH94} and assuming a constant gas to dust ratio of 100 and dust with a $\beta$ of 1.75 over the entire Galactic center , yielding a frequency-dependent dust opacity of $\kappa_\nu = \kappa_0\left(\frac{\nu}{\nu_0}\right)^\beta$, where $\kappa_0$ is
4.0 cm$^{2}$g$^{-1}$  at 505 GHz \citep[the method is described in more detail in ][]{Battersby11}. Toward the Southwest region, we measure an \htwo\, column of $4.9\times10^{22}$ cm$^{-2}$ in the ISO aperture, and toward the Northeast region, we measure an \htwo\, column of  $4.0\times10^{22}$ cm$^{-2}$, both consistent with the values shown in the column density maps of \cite{Etx11}. These column densities correspond to total molecular gas masses of 456 M$_{\odot}$  in the Southwest region and 370 M$_{\odot}$ in the northeast region. 

If our power-law extrapolation is valid, and if 50 K is a correct cutoff temperature for the CND, then the total molecular mass we infer from this method for the Northeast (Southwest) region would then account for 140 (32)\% of the total molecular mass inferred from the dust emission, or 125 (26)\% of the total molecular mass inferred from the CO observations. In contrast, the column of hot (T$>$500 K) \htwo\, measured from our discrete temperature fits toward the Northeast (Southwest) region only accounts for 6 (2)\% of the total \htwo\, column inferred from the dust emission, and 5 (2)\% of the total molecular mass inferred from the CO observations.

We find that dust and CO yield relatively consistent estimates of the total molecular gas within the ISO aperture, however, the molecular gas masses extrapolated from our continuous temperature fits to \htwo\ deviate significantly. In the Southwest region, our mass estimate from \htwo\, is only one quarter to one third of the total molecular gas measured from CO and dust. In contrast, in the Northeast region,  the mass estimate from \htwo\, is significantly larger than the total molecular gas measured from CO and dust. This extreme variation is largely due to the fact that the Southwest region has fainter \htwo\, and a correspondingly lower column density than the Northeast region, but yet is significantly brighter in CO and dust (and thus has a higher inferred total column of \htwo.) The discrepancy we find suggests that extrapolations of a power law temperature fit being used to estimate total molecular gas masses in extragalactic sources should be treated with caution, without a good understanding of where the extrapolation may become invalid. We further discuss several potential sources of this discrepancy in Section \ref{interpret}.

\section{Discussion}
\label{dis}

\subsection{The CND compared to other sources}
\label{compare}

To understand how the temperature of the warm \htwo\, and the fraction of gas at these temperatures in the CND match up with the properties of other sources, we first compare the CND to observations of other Galactic center molecular clouds. The pure-rotational 0-0 S(0), S(1), S(3), S(4) and S(5) lines of \htwo\, were observed in a sample of 16 clouds in the central 500 pc of the Galaxy by \cite{RF01}. As they used the \cite{Lutz} law to correct for the extinction of their sample, we have taken their reported line fluxes and redone the extinction correction using the \cite{Fritz11} law in order to be more consistent with our CND analysis. We select a subset of 5 clouds from their sample, all of which have detections of the 0-0 S(4) and S(5) lines. Fitting a power law temperature distribution to these clouds as we did for the CND, we find power law indices that range from 4.7 to 5.0, with fits shown in Figure \ref{fig10}. These are steeper than the power law indices we measure in the Northeast (n = 3.22) and Southwest (n = 2.83) regions of the CND, and indicate that the \htwo\, in typical Galactic center clouds is cooler than in the CND. This is consistent with a comparison of the mean T$_{76}$ for the 5 of these clouds for which the S(5) line could be detected \citep[700 K;][]{RF01} with what we would measure for T$_{76}$ from our CND data: 800 K for the Southwest region and 870 K for the Northeast region.

Similar to what we find when comparing to other molecular clouds in the central 500 pc of our Galaxy, the power law index for the temperature distribution in the Southwest and Northeast regions of CND (about 3) is also flatter than for any galaxies in the sample measured by \cite{Togi16}, which includes galaxies with star-forming nuclei, luminous and ultraluminous infrared galaxies (LIRGS and ULIRGS), Low Ionization Nuclear Emission Regions (LINERs), Seyfert Galaxies, dwarf galaxies, and radio galaxies. Measured power-law indices for this sample ranged from 3.79-6.4 with an average value of $\sim$ 4.84. The star-forming galaxies selected from the SINGS sample \citep{Kennicutt03} had a slightly lower average power law index of 4.5, but still do not approach the low values that we measure in the Circumnuclear disk. Even in the extreme shocked regions of Stephan's Quintet, the power law index is only measured to be $\sim$4.2 \citep{Appleton17}.

A direct comparison of these values is complicated because of the different scales involved: the CND comprises only a few parsecs in the center of our galaxy, while many of the measurements from \cite{Togi16} average over entire galaxies. Of course, the \htwo-emitting region of a galaxy may be much more compact: in the sample of ULIRGS, the mid-infrared \htwo\, emission is typically concentrated in the central 1 to a few kpc \citep{Higdon06}. Differences therefore could be either because even in galaxies more extreme than the milky way, conditions similar to the CND of our galaxy are not present over larger (hundreds of parsecs) scales, or because such conditions are present, but are diluted by abundant \htwo\, in less-extreme conditions at larger radii. In support of the latter scenario, we note that the `typical' power law indices we measure for clouds in the central 500 pc of our Galaxy (a region known to have higher temperatures, density, and turbulence than the disk of our Galaxy) appear quite comparable to those on kiloparsec scales in the centers of more extreme star-forming and infrared-bright galaxies. 

\subsection{The heating mechanism for the CND}
\label{heat}

\subsubsection{The excitation mechanism of \htwo\, in the CND}
For highly-excited lines of \htwo\, like those we analyze here, there are two possibilities for their excitation: collisional excitation, in which the observed levels are populated by collisions with other \htwo\, molecules, atoms, or electrons in energetic shocks \citep[e.g.,][]{Shull82}, and fluorescent excitation, in which the observed levels are populated by decay into lower states after the resonant absorption of ultraviolet photons in the Lyman and Werner bands \citep[e.g.,][]{Gould63,Black76}. Collisional excitation can be thought of as a ``bottom-up" population of the levels of \htwo\,, as persistent collisions in shocks with temperatures of thousands of K thermally excite \htwo\, out of its ground state and into the observed rotationally- and vibrationally-excited states. In contrast, fluorescent excitation can be thought of as a ``top-down" population of the levels of \htwo, as the molecules are directly excited into high vibrational levels by the resonant absorption of ultraviolet wavelength photons, and a fraction of the molecules, instead of dissociating, de-excite into the observed lower-excitation rotation and vibration states. A number of observational diagnostics exist to distinguish between these two excitation mechanisms, which focus on the relative brightness of rotational and highly vibrationally-excited transitions \citep[the latter of which are expected to be weaker or absent in the case of shock-excited emission;][]{Shull78,Black87,Wolfire91}.

Although fluorescent emission might seem to be ruled out due to the large radii at which the `very hot' \htwo\, traced by the 1-0 Q(1) line is observed, UV radiation from the central star cluster is not the only possible source of this excitation: cosmic rays, which as they impact the dense gas can also excite the Lyman and Werner bands of H$_2$ \citep{Prasad83}. Due to the limited sensitivity and wavelength range of the SWS spectra, we lack detections of many of these traditional tracers of fluorescent emission \citep[e.g., more highly vibrationally-excited lines like the 1-0 S(1) line at 2.1217 $\mu$m and the 2-1 S(1) line at  at 2.2476 $\mu$m][]{Shull78}. However, \cite{Tanaka89} previously observed these diagnostic lines toward both the Southwest and Northeast lobes of the CND, using the UKIRT telescope. They find that these sources, referred to as Sgr A-SW and NE, respectively, both have line ratios consistent with purely thermal excitation. Similarly, \cite{Ciurlo16} measure the 1-0 S(1) to 2-1 S(1) ratio in the central cavity of the CND (the central 36$''$ by 29$''$) and find regions where the excitation is dominated by both fluorescent and collisional excitation. However in the region that is likely most similar to the CND gas probed by the ISO observations (their 'Zone 1' on the inner edge of the CND, with a far stronger 1-0 S(1) line flux than in other locations analyzed), they find that the line ratios are consistent with collisional heating alone. We thus conclude that the \htwo\, lines we observe toward the Northeast and Southwest regions of the CND are collisionally excited. 

We also search for anomalous ortho to para ratios in our observed \htwo\, lines. The typical signature of disequilibrium in the ortho-para ratio is a zigzag or saw-tooth pattern in the Boltzmann diagram, where the ortho (odd $J$) lines appear systematically weaker than lines from the para (even $J$) transitions. We do see a slight sawtooth pattern in the S(1) through S(5) lines from the Southwest region, however the magnitude of the variation in these lines is consistent with the random scatter seen in the higher pure-rotational lines. We find no evidence for an anomalous ortho to para ratio in the observed pure-rotational lines of \htwo\, for the other two positions: this characteristic pattern does not appear in the residuals of the temperature fits for the Northeast region or toward Sgr A*. Anomalous ratios have previously been seen in the pure-rotational 0-0 S lines of several Galactic center clouds \citep{RF00}, as well as in the 1-0 S lines observed by \cite{Ciurlo16} in several positions in the central cavity of the CND. Such ratios have been attributed to the recent formation and/or destruction of \htwo\, before it can reach ortho to para equilibrium via e.g., proton-exchange collisions with H$^+$, H$_3^+$, or H$_3$O$^+$\citep{Gerlich90,LeB99}.

\subsubsection{Shock heating and line cooling}
\label{shock}
Based on the continuous distribution of warm to very hot temperatures we measure with the pure rotational lines of \htwo\, and the line ratios measured by \cite{Ciurlo16}, we favor a scenario in which the warm/hot \htwo\, in the CND is heated in shocks. The presence of gas at temperatures in excess of 2000 K likely rules out dissociative shocks as the primary source of this heating, as the \htwo\, molecules must survive to the post-shock phase in order to reach these temperatures, as opposed to dissociating and reforming at cooler temperatures \citep[$\sim$500 K;][]{Hollenbach89}.

Although either shocks or a high cosmic ray ionization rate are favored by chemical modeling of millimeter and submillimeter molecular lines in the CND \citep{Harada15}, heating by cosmic rays appears unlikely to be sufficient to produce the observed fractions of gas at these high temperatures \citep{Clark13}. This is consistent with the observed Herschel far-infrared line ratios measured by \cite{Goic13} toward the central cavity, which are also suggested to rule out both cosmic rays and X-rays for heating the hot (T$\sim$ 1300 K) gas. Photon-dominated regions (PDRs) as the sole source of heating in the CND were initially ruled out by \cite{Bradford05}, based on the observed ratio of CO 7-6 and [OI] 63 $\mu$m lines compared to predictions from PDR models \citep[e.g.,][]{Kaufman99}. However, this conclusion was based on the assumption that a clumpy PDR model with high-density gas like that applied to the Orion bar \citep[e.g.,][]{Burton90} is not applicable to the CND, when in fact more recent analyses have found that there is gas with significantly higher densities \citep[n$\sim2\times10^5-3\times10^6$ \cm;][]{RT12,Mills13} than the \cite{Bradford05} measurement (n$\sim5-6\times10^{4}$ \cm). The presence of higher-density clumps could allow for a more sizable contribution from PDR heating, however given that the CO 7-6 to [OI] 63 $\mu$m ratio in the CND is still an order of magnitude below a clumpy classical PDR like the Orion Bar \citep{Stacey93}, and that apparently very hot gas is also located at large separations from the central star cluster, it is still not likely that this is the sole source of heating in the CND. Some PDR heating could be supported by our measurement that the Southwest region (the edge of which is ionized) is warmer than the Northeast region. However, higher spatial resolution observations of the \htwo\, emission in conjunction with an improved 3D model of the molecular gas in the CND are needed to determine whether this difference is consistent with a radial trend in heating that would be expected for a PDR driven by the central star cluster.  

While the post-shock line emission resulting from a single planar C-shock model is well approximated as arising from an isothermal temperature distribution \citep[e.g.,][]{KN96,Neufeld06}, C-shocks (even single C-shocks) can also match power-law temperature distributions like those we observe. For example, single C-type bow shock models yield a temperature distribution with a power law index $\beta$ of 3.8 \citep[e.g.,][]{NY08}. Different values of $\beta$ can originate from different properties of single shocks \citep[e.g., a bow shock with a less curved structure should yield lower $\beta$;][]{YN11} or multiple shocks \citep[e.g., a mix of shock geometries, or a range of pre-shock densities, or a mix of shock velocities in which some shocks are sufficiently energetic to dissociate H$_2$][]{YN11,Appleton17,NY08}. Single bow shock models are traditionally applied to stellar jets and outflows, and are likely less applicable to the complex geometry of CND gas. However, a range of shock velocities (given the large turbulent linewidths) or pre-shock densities \citep[e.g., multiple density components are measured by][]{RT12} are both plausible scenarios for the CND. 

C-shocks as the dominant mode of heating the CND would be consistent with \cite{RF04}, who find that PDRs can only contribute 10-30\% of the heating for gas in Galactic center clouds outside of the CND. They favor moderate-velocity shocks (v$\sim$25 \kms) induced by turbulent motions as the primary heating source in these clouds, for example contributing to the warm \htwo\, measured by \cite{RF01} toward a number of these sources. For the CND, chemical modeling by \cite{Harada15} favors shocks with v$>$40 \kms\, for reproducing the observed abundances of species in millimeter and submillimeter observations. However, such velocities are near those required to dissociate \htwo\, molecules, and so velocities $\gtrsim$ 40 \kms\, are likely ruled out by our observations of \htwo\, at temperatures $>$ 2600 K, which are also consistent with what is found for the peak temperatures in C-shock models with v$\sim$25-30 \kms. Shock velocities $>$ 25 \kms\, and correspondingly higher temperatures in the CND would be consistent with recent work that suggests turbulent heating is responsible for a correlation between temperature and linewidth in Galactic center molecular clouds \citep{Immer16}, as the CND has much broader linewidths than other Galactic center clouds \citep[$\sigma \sim$10-40 \kms on ~0.2 pc scales;][]{MCHH09}, compared to predicted and observed $\sigma \sim$ 0.5-5 \kms for other Galactic center clouds \citep{Shetty12,Kauffmann13,Rathborne15}. In order to explain the higher temperatures measured in the Southwest, we would predict that careful analysis of the Northeast and Southwest regions would find larger linewidths in the Southwest, and more signatures of shocks.

In addition to providing insight into the gas heating mechanisms, the observed H$_2$ lines also make an important contribution to cooling the CND gas. Summing together the extinction-corrected fluxes of all of the observed pure rotational lines of H$_2$, and again using our adopted distance of 8 kpc, we measure a total H$_2$ luminosity of 275 L$_\odot$ toward the Northeast region, 200 L$_\odot$ toward the Southwest region, and 100 L$_\odot$ toward Sgr A* and the central cavity. The {\bf CND} luminosities can be compared to CO luminosities from \cite{Bradford05}, who conduct LVG modeling based on CO observations, and estimate a CO luminosity to mass ratio of 1.4 L$_\odot$ / M$_\odot$ for the entire CND. If we apply this to the CO masses adapted from \cite{RT12} in Table \ref{masses}, this would equate to total CO luminosities of roughly 580 L$_\odot$ toward the Northeast region and 780 L$_\odot$ toward the Southwest region. The H$_2$ lines would then contribute roughly 50\% and 25\% of the cooling generated by the CO line in the Northeast and Southwest regions, respectively.  For the CO conditions modeled by \citealt{Bradford05} (a uniform temp of 240 K), the authors predicted that the H$_2$ luminosity would be 30\% of CO luminosity, which is similar to the contribution we measure, though apparently originating in hotter gas. {\bf In the central cavity however, H$_2$ appears to provide a more significant fraction of the molecular gas cooling. Using observations of the far-infrared CO lines, \cite{Goic13} measure L$_{\mathrm{CO}}$= 125 L$_\odot$ toward the central $30''\times30''$. This is roughly three times the area of the aperture of the ISO observations, for which we measure L$_{\mathrm{H}2}$ = 100 L$_\odot$ toward the central cavity, indicating that H$_2$ is likely the dominant cooling line for molecular gas interior to the CND. However, this is still only a small fraction of the cooling  from the atomic gas in the central cavity through far-infrared fine structure lines \citep[e.g., L$_{\mathrm{[OIII]}}$ = 885 L$_\odot$, L$_{\mathrm{[OI]}}$ = 885 L$_\odot$, and  L$_{\mathrm{[CII]}}$ = 230 L$_\odot$ in a $30''\times30''$ aperture;][]{Goic13}.  }  

\subsection{The spatial distribution of the warm gas}
\label{spatial}
Given that the ISO spectra analyzed here consist of three large-aperture pointings toward the central parsecs and thus have extremely limited spatial information, one question is where the measured \htwo\, is actually located. Prior observations \citep[e.g.,][]{Etx11,Goic13,Lau13,Ciurlo16} have pointed to the central cavity and the inner edge of the CND as the location of both the hottest gas and dust. However, the resolution of our ISO data is not sufficient to distinguish between these locations, nor are the kinematics sufficiently precise to be indicative of a particular location \citep[though, the average velocities of the ISO Northwest and Southwest region spectra are broadly consistent with the kinematics of dense gas in the CND, e.g.,][]{Chris05}. However, the high-resolution maps of the 1-0  Q(1) line published in \cite{Feldmeier14} and reproduced here in Figure \ref{fig1} can give some insight into the distribution of the hottest \htwo\, in the central parsecs. We can clearly see peaks in the flux of the 1-0  Q(1) line at the positions of our ISO apertures centered on the Northeast and Southwest region. Importantly, even in this highly-excited \htwo\, line (which, as noted above is consistent with being thermalized at T $< 2600$ K), these peaks in the highest column density of the very hot \htwo\, are nearly cospatial with the peaks of cooler molecular gas and dust, as can be seen in a comparison with a map of HCN 4-3 from \cite{MCHH09} in Figure \ref{fig11}. In fact, emission from the 1-0 Q(1) line is clearly not confined just to the central cavity or a narrow inner edge of the CND, but is present at radii from 1 pc to 5 pc from Sgr A* and the central nuclear cluster.  Further, as can be seen in Figure \ref{fig11}, it is not simply the spatial distributions of H$_2$ and HCN which appear similar, but the gas kinematics as well.  So, although prior observations of hot molecular gas and dust in the central parsecs have suggested that this gas is primarily found in the central cavity and minispiral \citep{Cotera99b,Goic13,Ciurlo16} and the inner edge of the CND \citep{Lau13,Ciurlo16}, the observations of \cite{Feldmeier14} suggest that it is more broadly distributed throughout the CND. This is important for understanding what is heating the gas, and we return to this point in Section \ref{heat}. 

As expected, the column of hot \htwo\, is not greatest toward Sgr A* \citep[in the central cavity, which is believed to be largely evacuated, containing only a few tens of solar masses of ionized gas and a few hundred solar masses of neutral gas;][]{Jackson93}. It is also not clear in these ISO observations whether we are even detecting central cavity gas in the pointing toward Sgr A*. Due to large noise and poor baseline shapes for this position, we only detect enough lines of \htwo\, to constrain a single $\sim$ 1110 K tempertaure component. The velocity of the \htwo\, lines toward this position (28$\pm$25 \kms) could be consistent with the velocity of gas in the nearby 50 and 20 kms clouds, which recent orbital models place $\gtrsim$ 50 pc in the foreground of Sgr A* \citep{Kruijssen15}. The spectra also do not clearly have a broader line width than the other positions, which would be expected if this was indeed gas closer to Sgr A*, with higher orbital speeds consistent with those seen in the ionized minispiral \citep{Zhao93}. So, although \cite{Ciurlo16} appear to detect gas in the central cavity in their higher-resolution maps of near-infrared rovibrational lines, we cannot clearly say that the gas we detect toward Sgr A* is any closer to the central black hole than the CND gas detected in other positions. 

Despite the lack of detailed spatial information in the ISO data, we do notice several key differences between the \htwo\, emission toward the Northeast and Southwest regions of the CND. First, the \htwo\, lines are stronger for a given $J$ in the Northeast region than in the Southwest region. Not surprisingly, the total column of warm \htwo\, in the Northeast region is twice as large as in the Southwest region. However, even though the lines are brighter in the Northeast region, the extinction toward the Northeast region is slightly larger. Interestingly, this is the reverse of what is typically observed in the emission from other molecular species, where the Northeast region generally appears much weaker than the Southwest region \citep{RT12,Mills13}. 

\subsection{Interpreting the mass of warm \htwo\, in the CND}
\label{interpret}

Using the measured power law temperature distributions to measure a total H$_2$ column by extrapolating to a common cutoff temperature of 50 K, we find that relative to estimates based on dust continuum emission and CO line emission we overestimate the molecular gas mass in Northeast region, and underestimate the mass in the Southwest region. This indicates that determining masses in this way is not straightforward, and may even not be valid for certain scenarios. In order to begin to assess the factors that determine whether the extrapolation of a power law temperature distribution yields a valid \htwo\, mass in the CND, we focus on several key assumptions of this analysis. The first is that T = 50 K (or T= 100 K)  is a valid cutoff temperature for the gas in the CND. A second major assumption is that the extrapolation of the power-law index derived using the S(1) lines and above is valid for the lower temperatures of interest here. A third assumption is that we are able to detect all of the emitting \htwo\, in our aperture (there are no local extinction effects). Finally, in order to put the measured quantities in context, we must assume that either CO or dust is a reliable indicator of the total \htwo\, column.

\subsubsection{The cutoff temperature}
\label{cutoff}

We first address the assumption that a temperature of 50 or 100 K is a valid cutoff temperature for the distribution of gas temperatures in the CND. First, in order to match the mass estimates from CO and the dust continuum, the cutoff temperature should be above 50 K in the Northeast region, and below 50 K in the Southwest region. Immediately, this appears inconsistent with the general trend seen that the Southwest region is warmer than the Northeast. However, we also want to examine whether there is any additional evidence that such cold gas could exist in this region. The coolest rotational temperature we can directly measure through our discrete temperature fits is 580 K in the Southwest region and 520 K in the Northeast region. The non-detection of the S(0) line suggests there may be an additional gas component with temperature $>$110 K ($>$130 K). However, this is not so strongly constraining as to rule out separate and yet cooler temperature components, including gas too cool to significantly excite the S(0) line. The best constraint on this cold material comes from the dust, as temperatures as low as 25 K have been measured for the CND \citep{Etx11}. Although dust and gas temperatures are not globally observed to be in thermal equilibrium in the Galactic center, dust temperatures of 20-30 are typically associated with gas temperatures of 50-70 K in other Galactic center clouds \citep{Ginsburg16}, suggesting that the presence of gas significantly cooler than 50 K is unlikely. 

A cutoff temperature of 50 K also implies that the bulk of the gas ($\sim70-80\%$) is cold (T$<$100 K). Although gas temperatures of 50 K are not excluded by excitation analyses of HCN \citep{Mills13}, the CO analysis of \cite{RT12} found the temperature of the CO must be greater than 150 K (based on the observed ratio of $^{12}$CO and $^{13}$CO lines). While this would appear to rule out cutoff temperatures as low as 50 or even 100 K, we note that without any cool gas, the total \htwo\, column we infer would be entirely inconsistent with the H$_2$ column inferred from these same CO observations. Either there is cooler gas present, or we are not detecting a large fraction of the H$_2$ emission from the CND. We thus do not yet rule out the possibility of a cutoff temperature as low as 50 K for the CND gas. However, this scenario then requires a significant burden of observational proof, including an explanation of the observed CO isotope ratios (either as an average on large scales that is not representative of the isotope ratios on smaller scales, or due to some type of anomaly), and verification of T= 50 or 100 K gas with direct measurements using other tracers (e.g. \am, CH$_3$CN, CH$_3$CCH).

\subsubsection{The extrapolation of a single power-law temperature fit}
In order to infer a total gas mass by extrapolating our \htwo\, data to temperatures below those which are measured, the power law temperature distribution we fit to the S(1) and higher lines must also be valid for the lower-temperature gas. In observations of the centers of normal spirals and star forming galaxies that include the detection of the S(0) line, there is no evidence for a break in the power law between the S(0) and S(1) lines \citep{Togi16}. Further, extrapolating the power-law distribution of \htwo\, column densities to a temperature of 50 K in these galaxies appears to consistently recover the total \htwo\, column as inferred through measurements of CO. This would support the adoption of a single power-law distribution for describing the range of temperatures present in the CND gas. Alternatively, if the gas in the CND were not consistent with a single power-law description of the temperature, this could account for the small (and varying; 6-140\%) fraction of the total (CO and dust-derived) \htwo\, column that is recovered in the Southwest and Northeast regions of the CND. However, a more complicated temperature profile would seem to require a unique physical mechanism in this region that does not operate on kiloparsec scales in the center of other galaxies. Although the local line widths and densities in CND may be slightly more extreme than the average conditions over kiloparsec scales in the center of our own and other galaxies \citep[e.g.][]{MCHH09,Mills13}, at present we do not have a good justification for invoking the additional complexity of a varying power-law index.

\subsubsection{Local variations in the extinction}
Another possible factor that could explain the fact that total \htwo\, column derived from our power-law extrapolation of the pure-rotational lines, and that derived from CO or dust are not completely consistent (especially in the Southwest region) is clumpy extinction. In this scenario, the lower column of pure-rotational line emission observed in the Southwest region \citep[where stronger emission from dense gas tracers is observed;][]{MCHH09} might be due to having more of the gas hidden by higher local extinction in the clumpy structures that carry the bulk of the mass. If extinction did hide some of the \htwo\, emission in the CND, this could relax the tension we found from comparison with the CO observations of \cite{RT12} that both require the \htwo\, to be hotter than 150 K, and infer a larger total \htwo\, column than can be matched with our direct observations of \htwo\, even when extrapolating to 50 K. A cutoff temperature of 150 K would then not be inconsistent with a total column inferred directly from \htwo\, that is much less than that derived with CO (or dust). 

We note that due to the flat shape of the extinction law from 2-20 $\mu$m (Fritz, and our plots of extinction), higher clumpy extinction for one region should not manifest as any change in temperature, as emission would be uniformly missing for all of the observed rotational lines. However, the fact that the extinction is actually measured to be slightly higher in the Northeast region (while it would need to be higher in the Southwest region to hide more of the \htwo\, emission) would seem to argue against the scenario we propose. Ultimately, higher spatial and spectral resolution observations are needed to directly compare the clumpy structure observed in other molecules with the structure of the \htwo\, emission, and to and compare the kinematic components directly traced by \htwo\, and other molecules to confirm that \htwo\, is not subject to significant local extinction, and that the \htwo\, is indeed observed to be fully and proportionately intermixed with the gas traced by other molecules. 

\subsubsection{Proxies for the total \htwo\, mass}

We find that the total molecular mass inferred from two proxies for \htwo: observations of multiple lines of $^{12}$CO and $^{13}$CO \citep{RT12} and column density maps derived from Herschel dust emission (Battersby et al. in prep; private communication) are generally consistent. The dust-derived masses are slightly lower than the CO-derived masses: by a factor of 1.2 in the Southwest and 1.1 in the Northeast. However, the CO-derived mass has been scaled to the size of the ISO aperture (as the aperture for the CO observations is significantly larger) assuming that the emission is uniform across this aperture. If this emission is not uniform, then there could be a larger discrepancy between these masses.  Given that the dust-derived mass is consistent with the CO-derived mass, and that the Galactic center consists of relatively high-metallicity gas, we also do not expect that CO-dark gas is a substantial component of the CND. 

While it is possible that there could be additional systematic uncertainties in these mass determinations (for example, in the adopted [$^{12}$CO]/[\htwo] abundance of $8\times10^{-5}$ used by \cite{RT12} to determine the CO to \htwo\, conversion, or in the assumption for the dust of a given dust mass opacity $\kappa$, a constant gas to dust ratio of 100 or a $\beta$ of 1.75 over the entire Galactic center), the consistency of the masses inferred from both CO and dust suggest that these are unlikely to be significant. 

\subsubsection{The total \htwo\, mass}

We can make an extremely rough estimate of the total molecular mass of the CND by using the maps of the Q(1) line flux from \cite{Feldmeier14},  to scale the masses determined from the rotational \htwo\, lines in the observed ISO apertures. The two 14$''\times20''$ ISO apertures covering the Northeast and Southwest regions contain only $\sim$1/13 of the total Q(1) emission seen by \cite{Feldmeier14} over the central 9.5$\times$8 parsecs of the CND. If the brightness distribution of the rotational lines we use to determine the mass and the rovibrational Q(1) line of \htwo\, is similar, then we would infer that the total mass of gas with T$>$ 50 K in the CND is $\sim$ 8600 M$_{\odot}$. This is $\sim$15-40\% of the total molecular mass of gas (2-5$\times10^4$ \msun) estimated to be present in the CND \citep{Etx11,RT12}. However, there are likely to be significant differences in the distribution of rotational and rovibrational lines, so this estimate should be treated with caution. We also note that the estimates of the total mass from CO and dust also suffer from limitations: the mass estimate from \cite{RT12} is extrapolated from LVG fits in just two apertures over a limited velocity range, and the estimate from \cite{Etx11}, while covering the bulk of the CND, is limited to a central radius of 1.75 pc.

\section{Conclusion}

We have analyzed archival ISO observations of the spectra of three positions toward the central parsecs of our Galaxy, pointed toward the central supermassive black hole Sgr A*, and the Northeast and Southwest regions of the CND. Below, we summarize the main findings of this paper:

\begin{itemize}

\item In both the Northwest and Southeast regions of the CND we detect emission from pure-rotational lines of \htwo\, up to the 0-0 S(13) lines, as well as from a handful of rovibrational transitions. 

\item From the pure-rotational lines of \htwo, we are able to fit three discrete temperature components to the hot gas toward both regions of the CND. We find slightly lower best-fit temperatures in the Northeast region (520 K, 1260 K, 2840 K ) than in the Southwest region (580 K, 1350 K, 3630 K) although in both cases the highest temperature present is loosely constrained, and is best interpreted as a lower limit of T $\gtrsim$ 2600 K.

\item From our discrete temperature fits, we measure a total column and mass of hot \htwo\, that is approximately twice as large for the Northeast region (N$_{\mathrm{H}2}$ = 1.20$\times10^{21}$ cm$^{-2}$; M$_{\mathrm{H}2}$ = 11.4 M$_{\odot}$ ) as the Southwest region (N$_{\mathrm{H}2}$ = 6.43$\times10^{20}$ cm$^{-2}$, M$_{\mathrm{H}2}$ = 6.1 M$_{\odot}$). Comparing these columns to the total column of \htwo\, estimated from CO and dust emission, we estimate that the warm/hot \htwo\, represents between 0.2-2\% of all of the \htwo\, in the CND. 

\item We also fit a continuous, power-law distribution of temperatures to the observed pure-rotational lines of \htwo. We find a power law index that is higher in the Northeast (3.22) than in the Southwest (2.83), confirming that the molecular gas is warmer in the Southwest region of the CND. These power law indices are also higher than those we measure for other clouds in the Galactic center (4.7-5.0), and higher than previously measured for other galaxies \citep[3.79-6.4;][]{Togi16}.

\item Extrapolating the power-law distribution we fit down to a temperature of 50 K, we would estimate a total mass of 146 M$_{\odot}$ for the Southwest region and 518 M$_{\odot}$ for the Northeast region. For the Southwest (Northeast) region, these masses represent 32 (140) \% of the total molecular mass inferred from measurements of the dust emission from Battersby et al. (in prep.; private communication), and 26 (125) \% of the total molecular mass inferred from measurements of the CO emission \citep{RT12}.

\item Based on the observed \htwo\, properties as well as prior work measuring the spatial distribution and line ratios of rovibrationally-excited \htwo\, we favor C-shocks as the primary excitation mechanism for the warm/hot \htwo\, and the dominant heating source for the CND gas. 

\end{itemize}

\section{Acknowledgements}
{\bf We especially thank the anonymous referee for their work in making extremely detailed and helpful comments that significantly improved the presentation of these results.}E.A.C.M thanks Nadine Neumayer for drawing attention to \htwo\, observations of the Circumnuclear disk, and Anya Feldmeier-Krause for sharing her 1-0 Q(1) data of the central 8 pc.  A.T.  is grateful for support from National Science Foundation under Grant No. 1616828

\bibliographystyle{hapj}
\bibliography{H2}

\clearpage
\section{Figures and Tables}

\clearpage

\clearpage

\begin{figure}[tbh]
\includegraphics[scale=0.5]{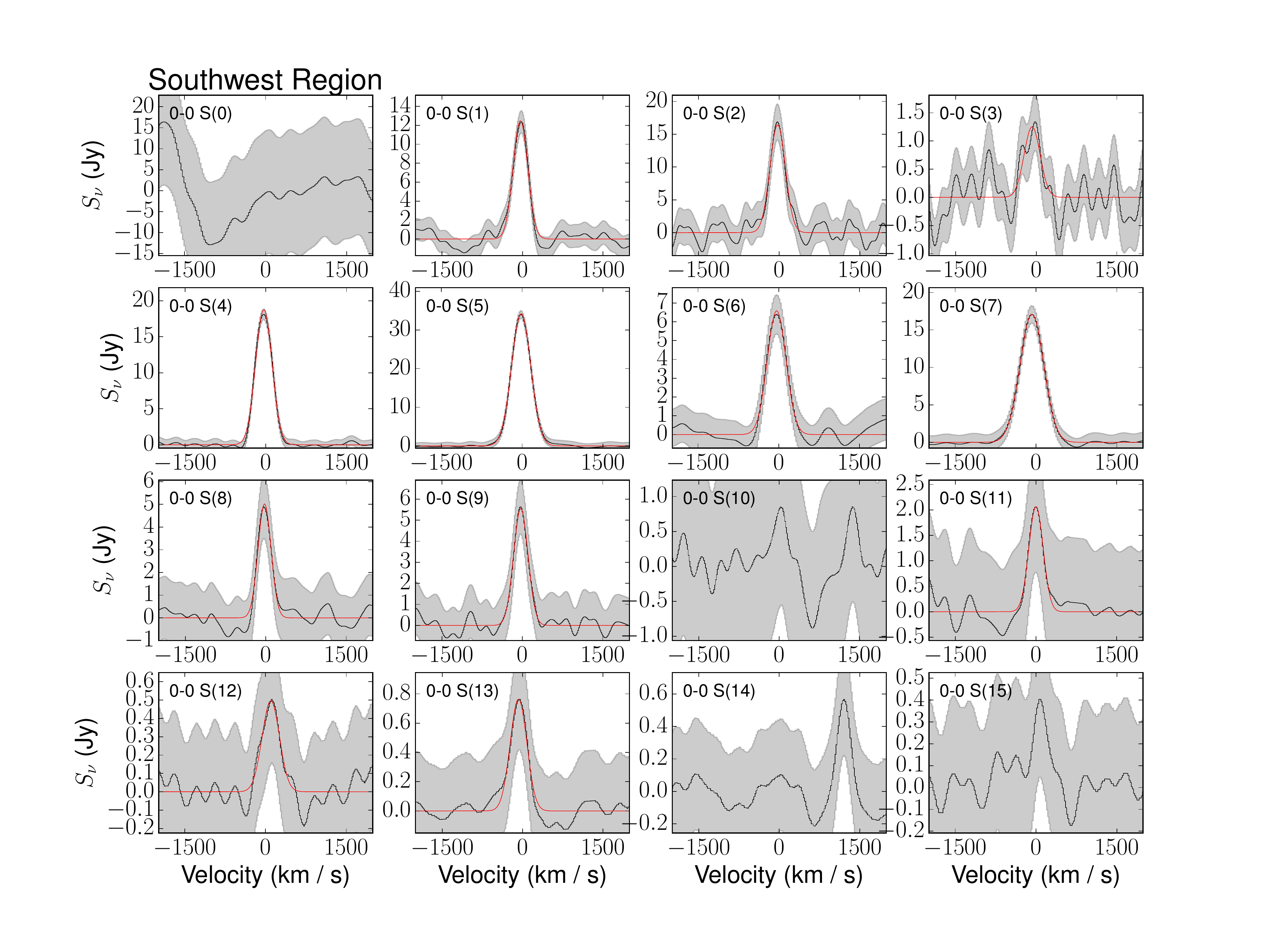}
\caption{Spectra of the pure-rotational lines of \htwo\, from 0-0 S(0) to S(15) toward the Southwest region, Northeast region, and Sgr A*. The amplitude uncertainty at each wavelength is shown in grey shading. Gaussian fits to all of the significantly-detected lines are overplotted in red. }
\vspace{-0.7cm}	
\label{fig3a}
\end{figure}
\clearpage

\renewcommand\thefigure{3}
\begin{figure}[tbh]
\includegraphics[scale=0.5]{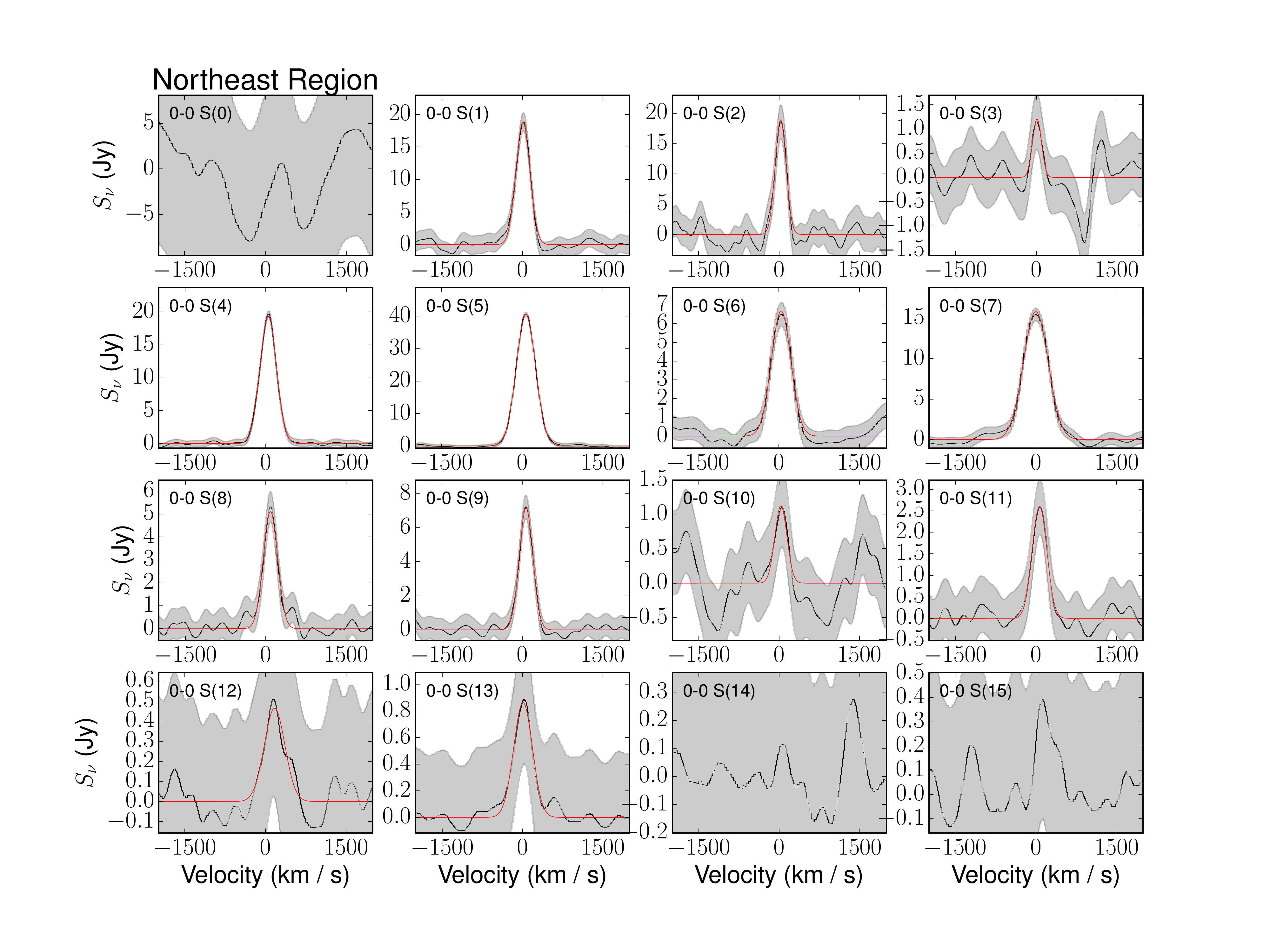}
\caption{Continued.}
\vspace{-0.7cm}	
\label{fig3b}
\end{figure}
\clearpage

\renewcommand\thefigure{3}
\begin{figure}[tbh]
\includegraphics[scale=0.5]{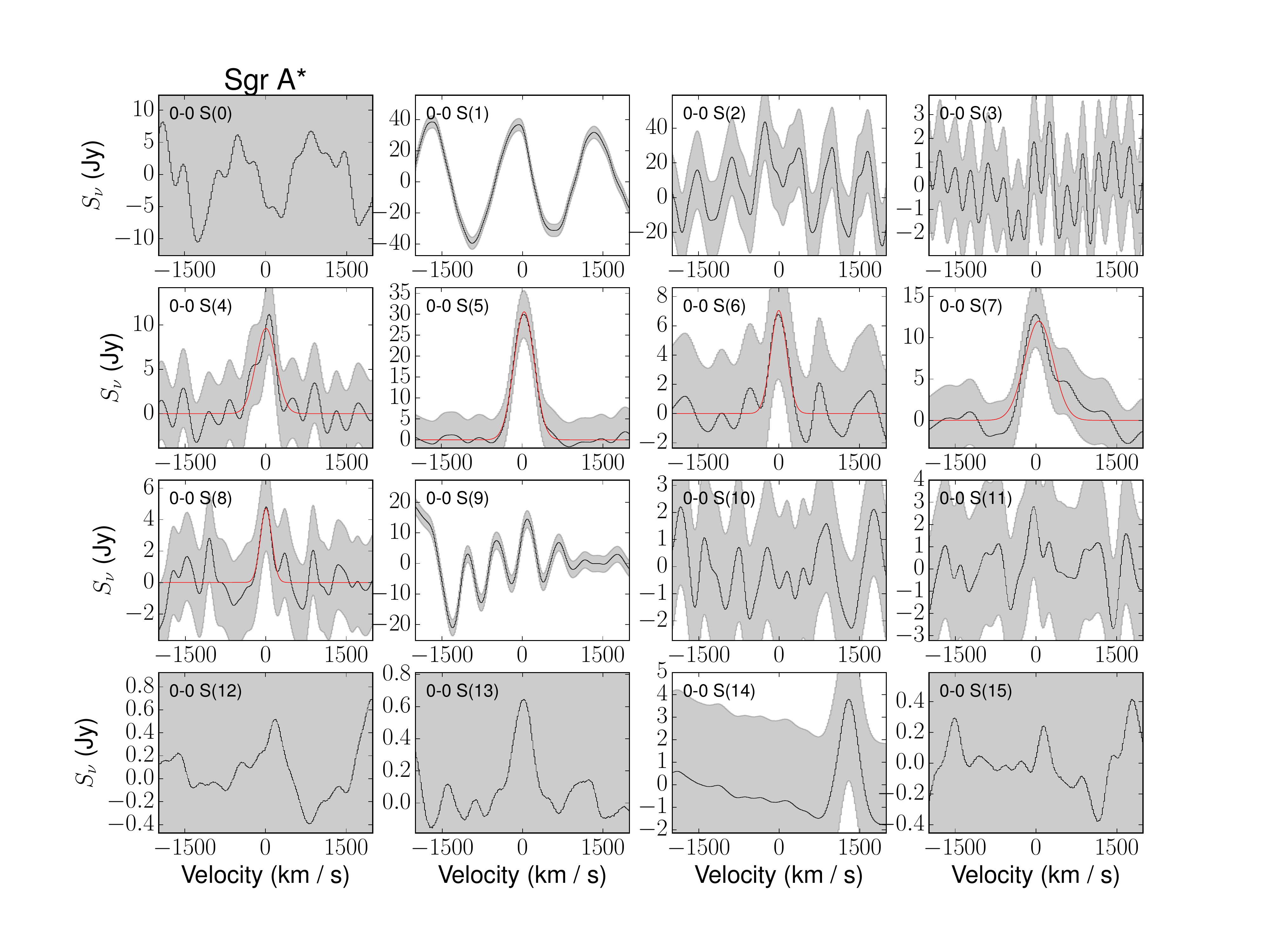}
\caption{Continued.}
\vspace{-0.7cm}	
\label{fig3c}
\end{figure}
\clearpage

\renewcommand\thefigure{4}
\begin{figure}[tbh]
\vspace{-1cm} 
\includegraphics[scale=0.5]{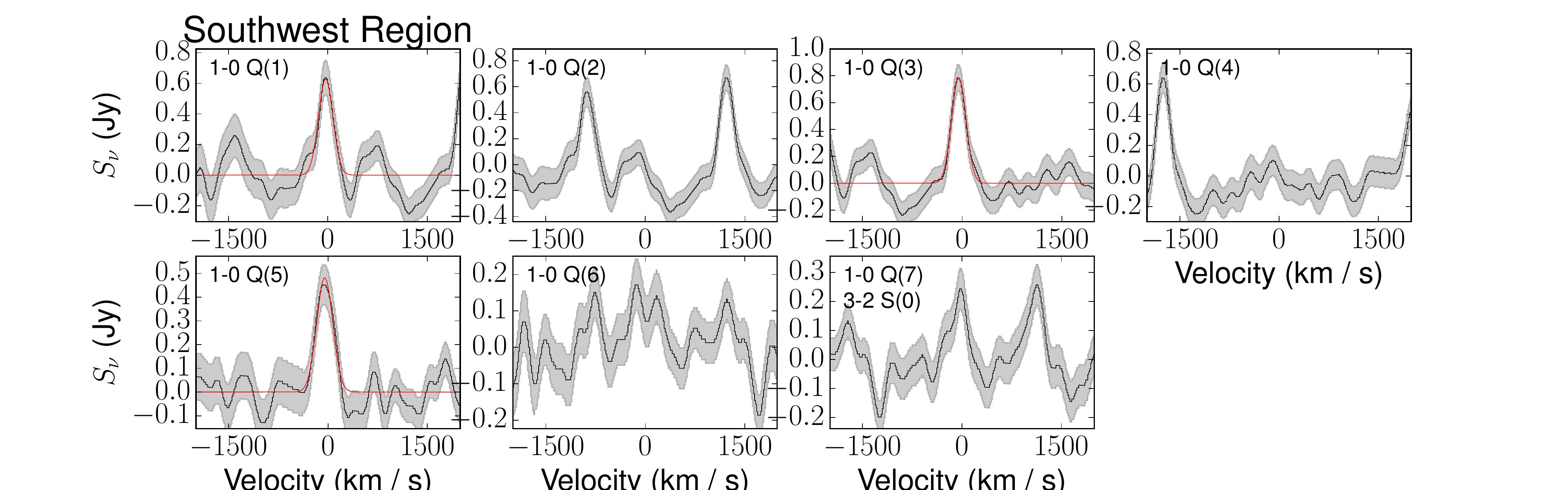}
\includegraphics[scale=0.5]{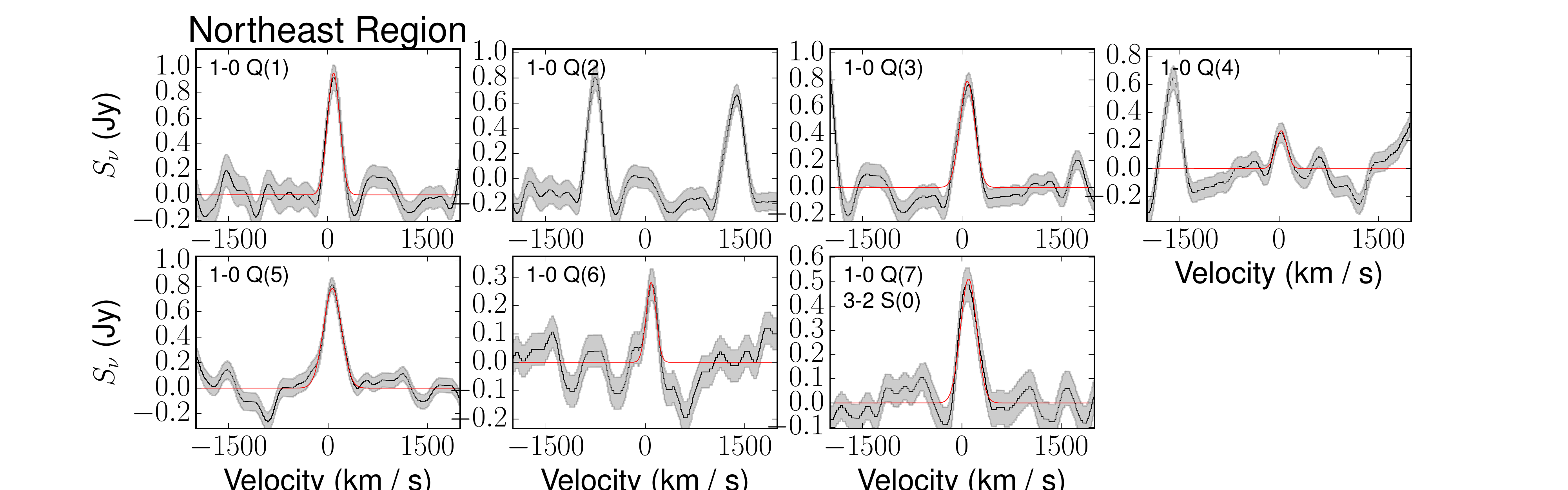}
\includegraphics[scale=0.5]{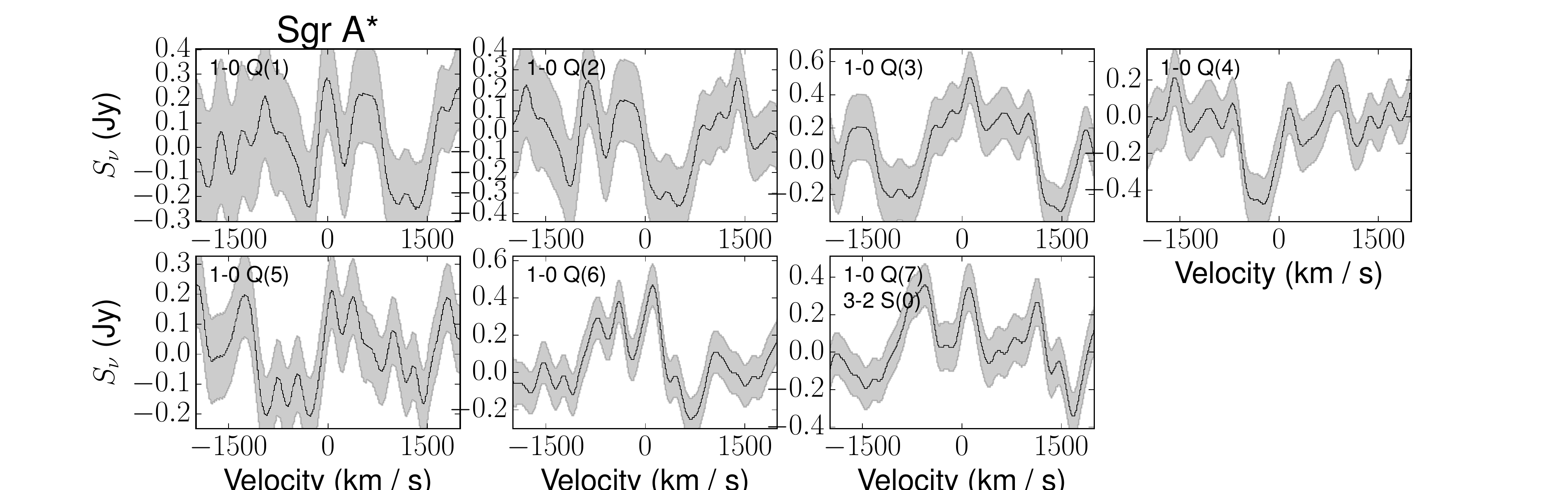}
\caption{Spectra of the 1-0 Q(1) though Q(7) lines toward the Southwest region, Northeast region, and Sgr A*. The amplitude uncertainty at each wavelength is shown in grey shading. Gaussian fits to all of the significantly-detected lines are overplotted in red. }
\vspace{-0.7cm}	
\label{fig4}
\end{figure}
\clearpage

\renewcommand\thefigure{5}
\begin{figure}[tbh]
\includegraphics[scale=0.5]{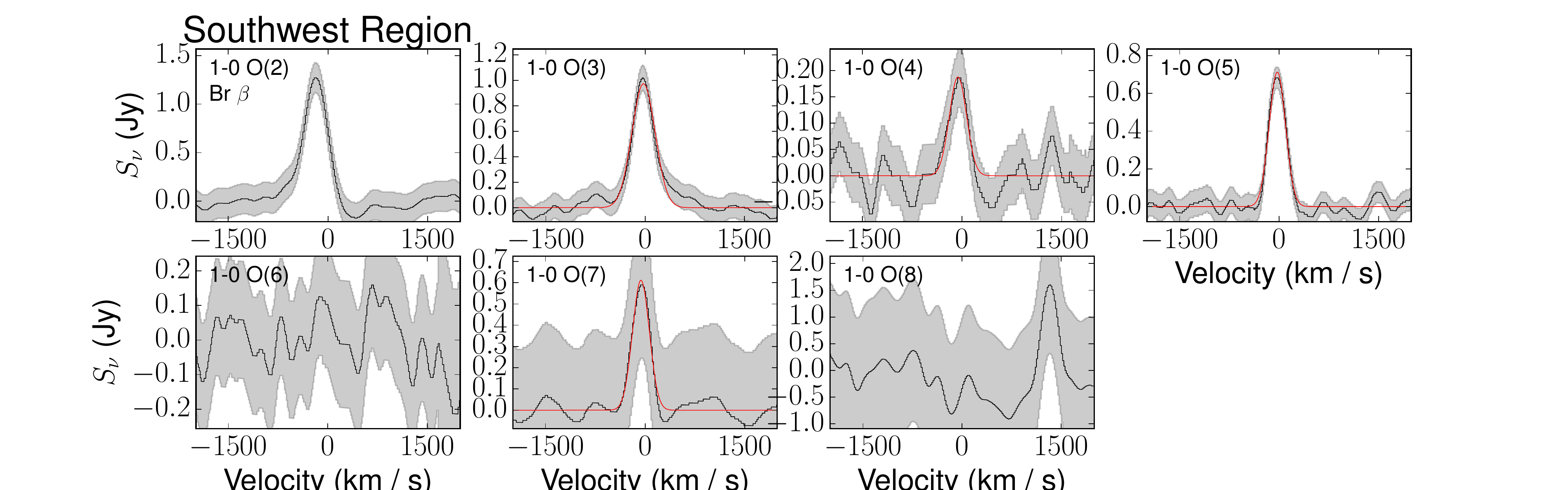}
\includegraphics[scale=0.5]{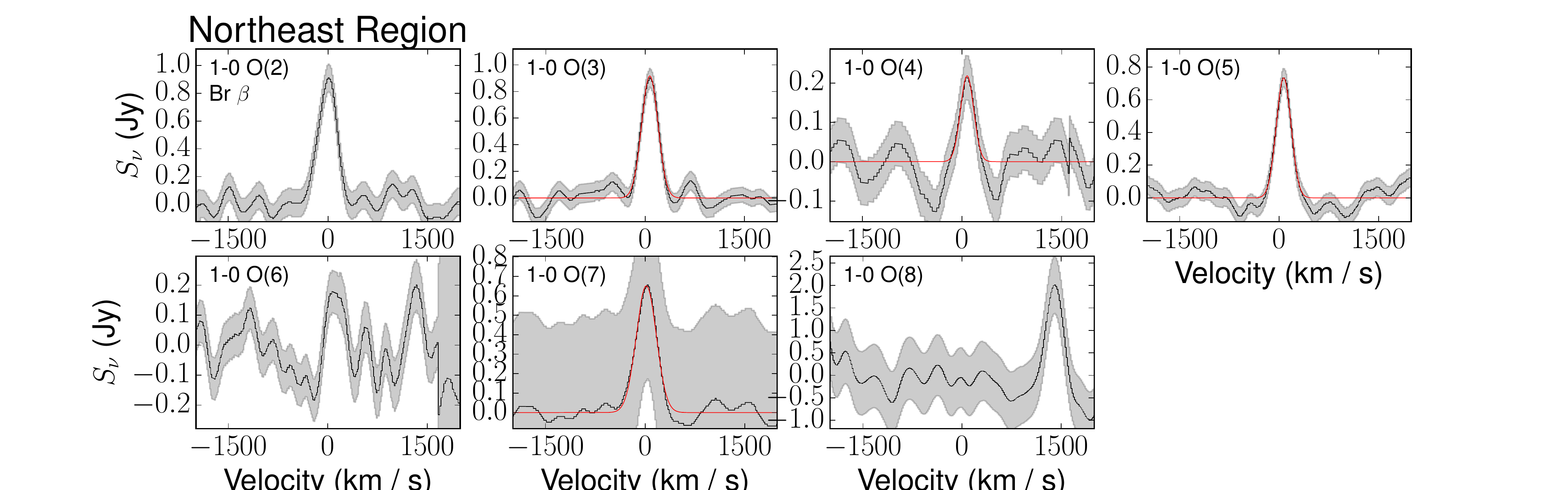}
\includegraphics[scale=0.5]{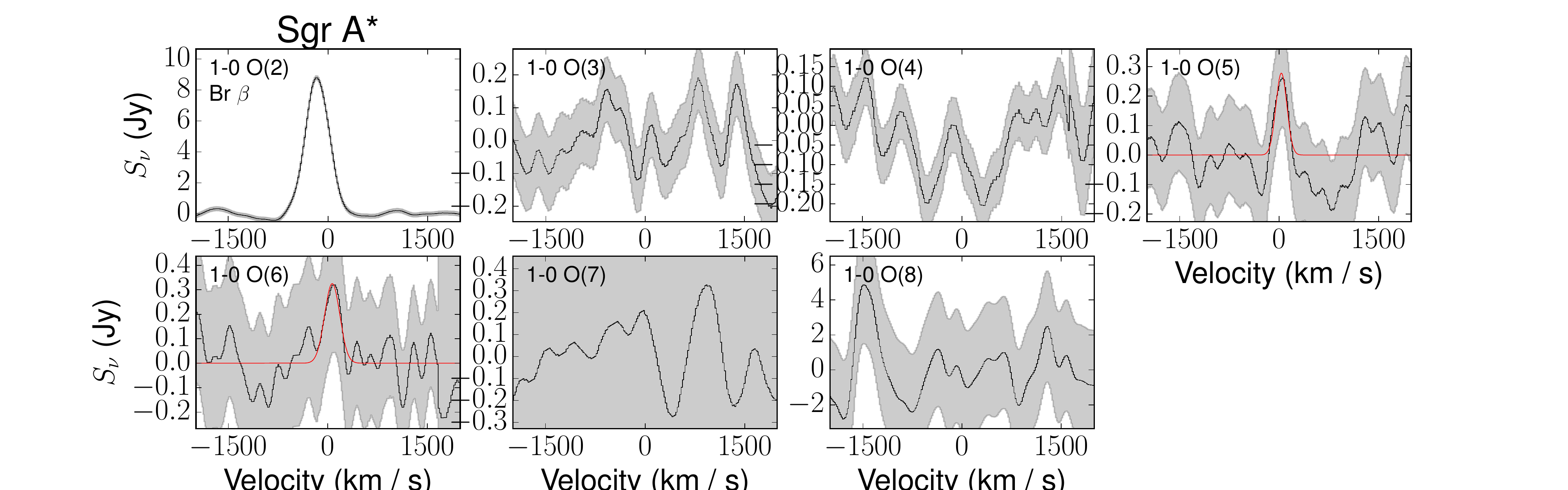}
\caption{Spectra of the 1-0 O(2) though O(8) lines toward the Southwest region, Northeast region, and Sgr A*. The amplitude uncertainty at each wavelength is shown in grey shading. Gaussian fits to all of the significantly-detected lines are overplotted in red. }
\vspace{-0.7cm}	
\label{fig5}
\end{figure}
\clearpage

\renewcommand\thefigure{6}
\begin{figure}[tbh]
\vspace{0.1cm}
\hspace{-0.1cm}
\includegraphics[scale=0.4]{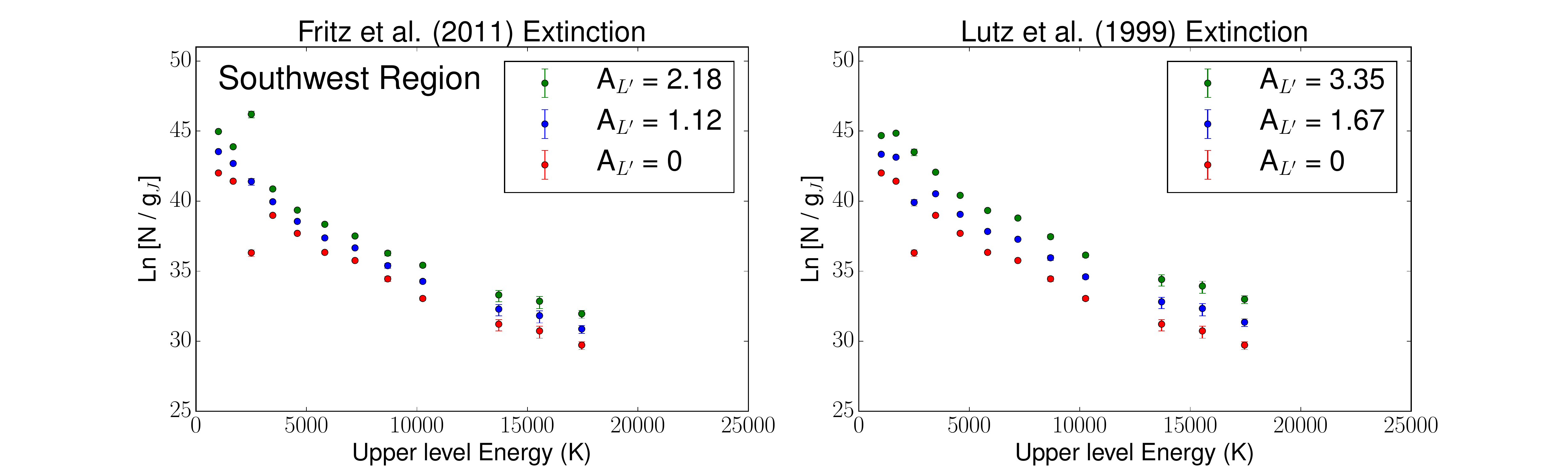}
\includegraphics[scale=0.4]{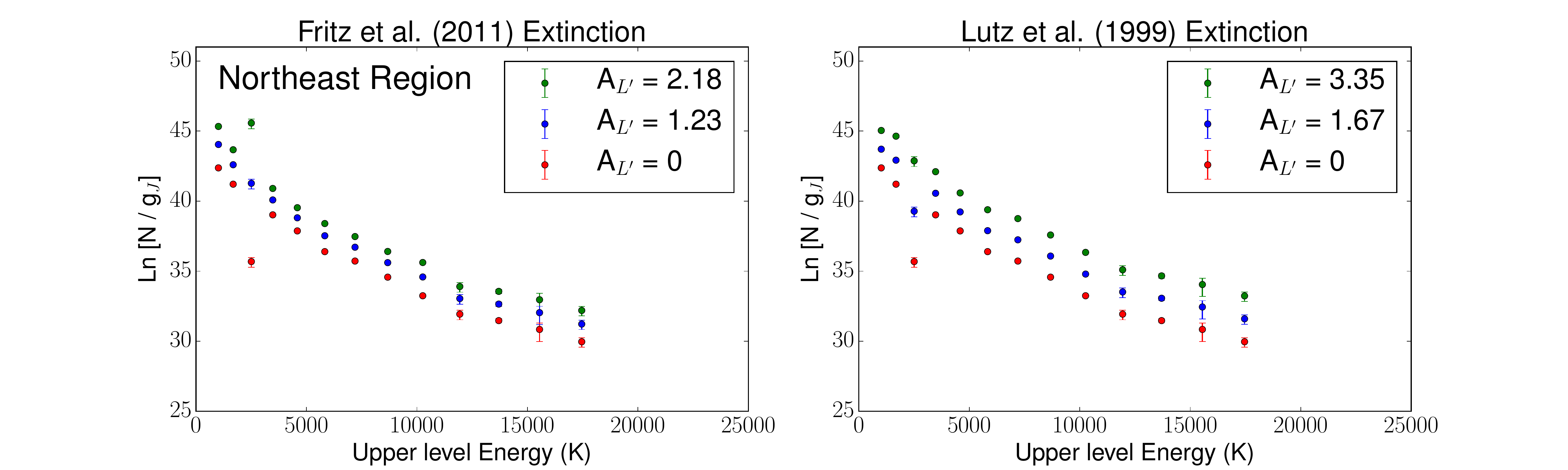}
\caption{Boltzmann plots of the 0-0 S lines toward the Southwest region and Northeast region of the CND. {\bf Left:}  Here, we apply the extinction law derived by \cite{Fritz11} from the ISO observations of Sgr A*. We show the column densities first uncorrected for extinction (red), then with an optimal correction fit in conjunction with the temperature components to minimize the residual scatter (blue), and finally for an overcorrection of the extinction (green). {\bf Right:} Here, we apply the extinction law derived by \cite{Lutz} from the ISO observations of Sgr A*. We show the column densities first uncorrected for extinction (red), then with a correction equivalent to 30 magnitudes of visual extinction (blue), and finally for the extinction correction needed to make the S(3) line consistent with neighboring lines (green).}
\vspace{-0.7cm}	
\label{fig6}
\end{figure}
\clearpage

\renewcommand\thefigure{7}
\begin{figure}[tbh]
\vspace{0.1cm}
\hspace{-0.1cm}
\includegraphics[scale=0.4]{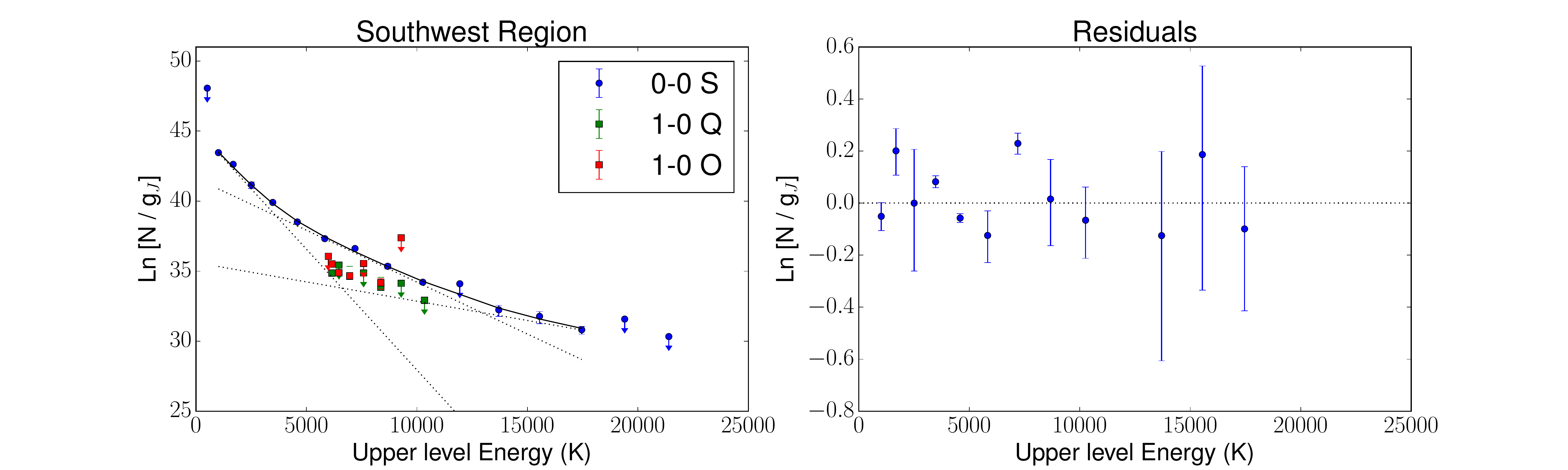}
\includegraphics[scale=0.4]{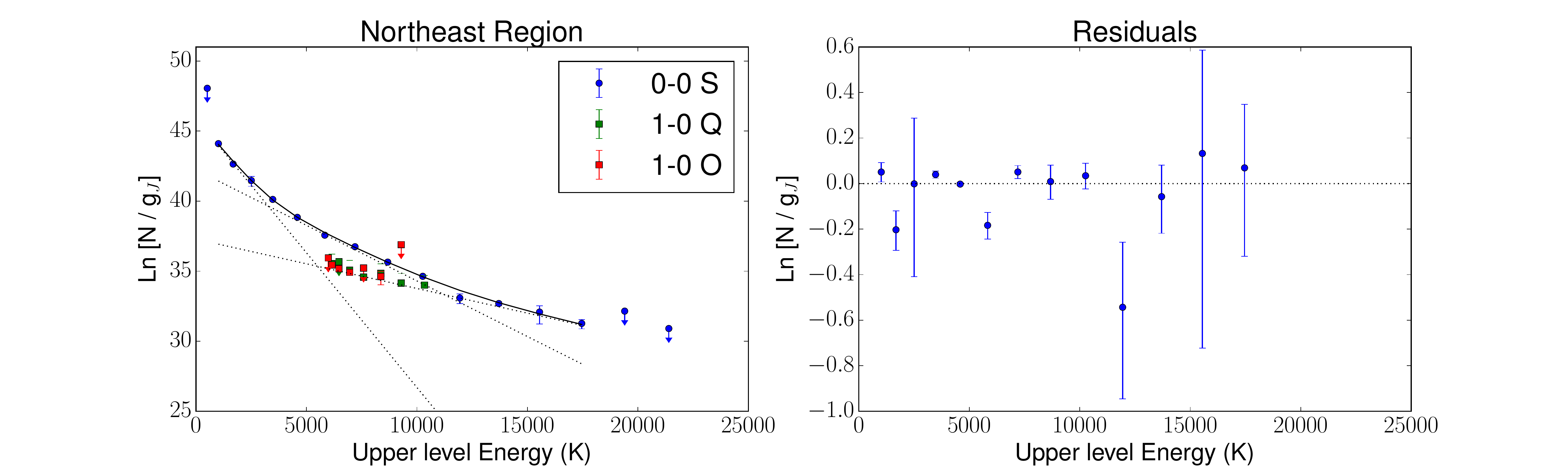}
\includegraphics[scale=0.4]{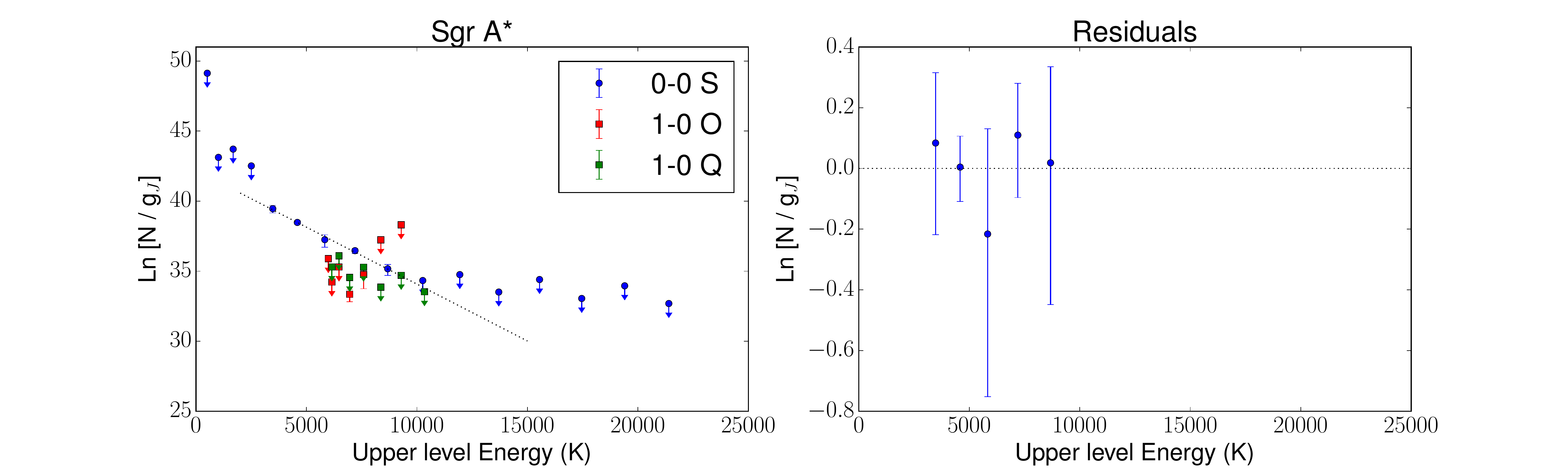}
\caption{Boltzmann plots of the observed \htwo\, lines (0-0 S, 1-0 Q, and 1-0 O) toward the Southwest region, Northeast region, and Sgr A*. The plotted column densities have been calculated from line fluxes that have been corrected for extinction.  Fits to the 0-0 S column densities for three discrete temperature components (dashed lines) and their summation (solid line) are shown for the Southwest and Northeast regions. The properties of these temperature components are given in Table \ref{fraction}. Only one temperature component (dashed line) can be fit toward Sgr A*. The residuals of the fits to the 0-0 S lines are also show in the lefthand panels.}
\vspace{-0.7cm}	
\label{fig7}
\end{figure}
\clearpage

\renewcommand\thefigure{8}
\begin{figure}[tbh]
\includegraphics[scale=0.5]{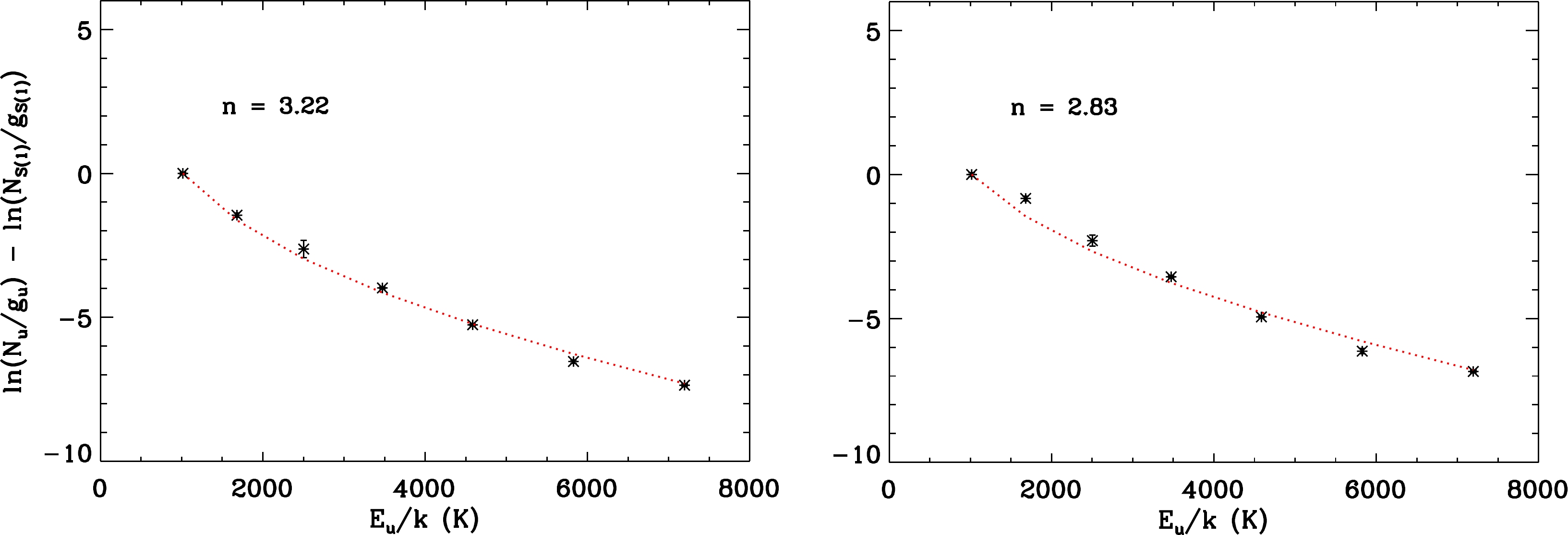}
\caption{Fits to the observed 0-0 S(1) through 0-0 S(7) \htwo\, lines for a power-law distribution of temperatures toward the Southwest (left) and Northeast (right) regions. The plotted column densities have been calculated from line fluxes that have been corrected for extinction.}
\vspace{-0.7cm}	
\label{fig8}
\end{figure}
\clearpage

\renewcommand\thefigure{9}
\begin{figure}[tbh]
\vspace{0.1cm}
\hspace{-0.1cm}
\includegraphics[scale=0.5]{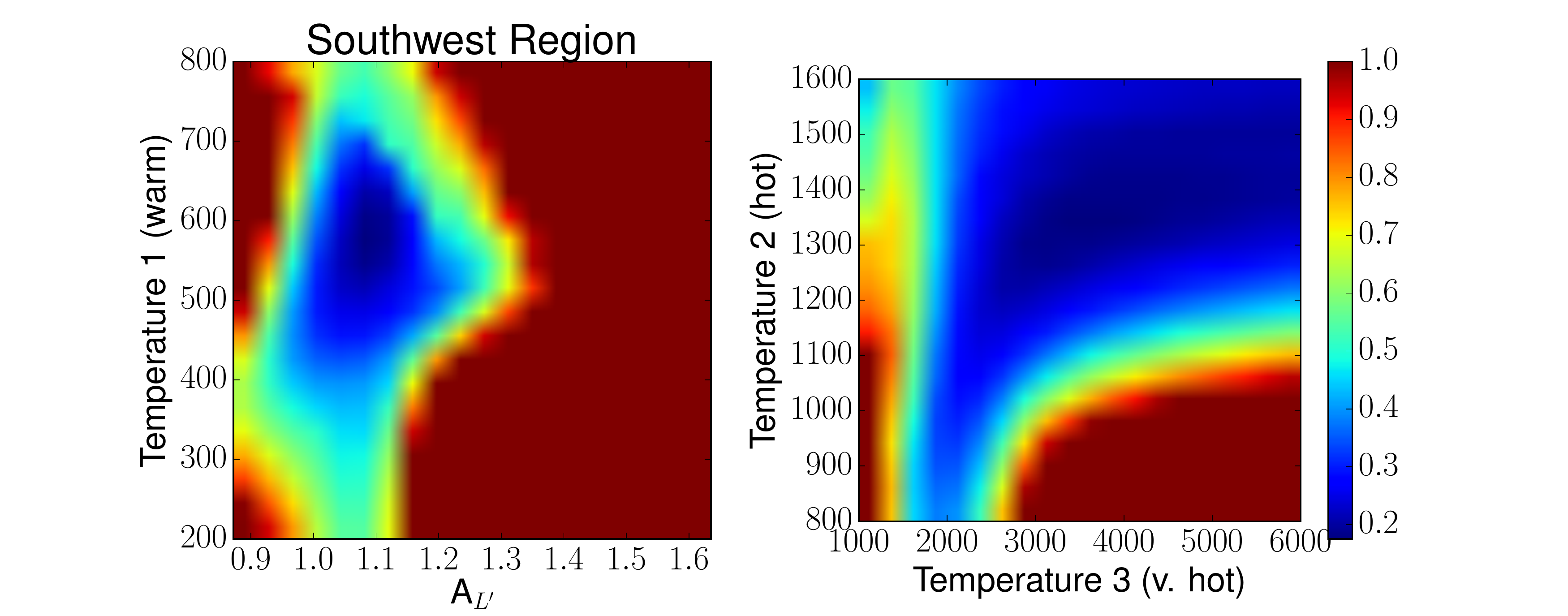}
\includegraphics[scale=0.5]{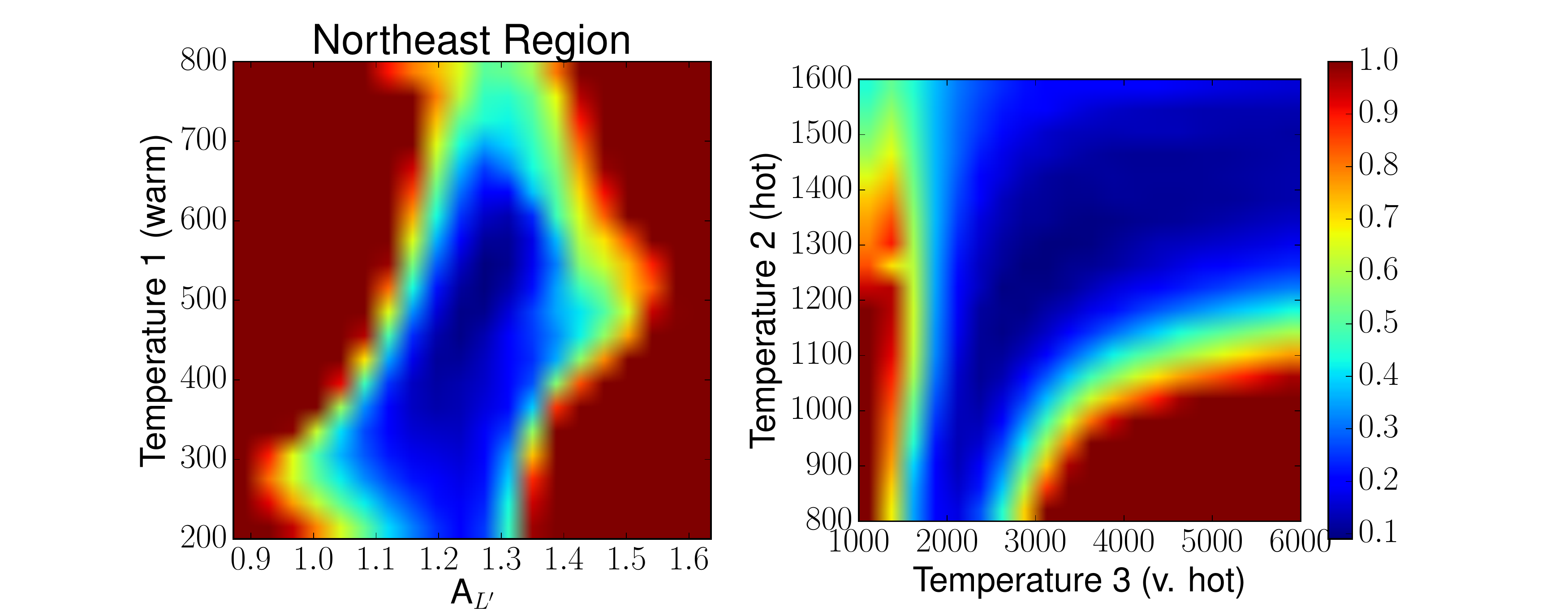}
\caption{Plots of the residual scatter from simultaneous fits to the L$'$ band ($\lambda$ = 3.776 $\mu$m) extinction and three temperature components, showing the constraints that can be achieved for these four parameters. The color scale in these plots represents the reduced $\chi$-squared value of the fit. In general, the L$'$-band extinction, warm temperature, and hot temperature are well constrained by these observations in both the Southwest and Northeast region. However, the very hot temperature is less well-constrained, and is best understood as a lower limit on this value.}
\vspace{-0.7cm}	
\label{fig9}
\end{figure}
\clearpage

\renewcommand\thefigure{10}
\begin{figure}[tbh]
\includegraphics[scale=0.5]{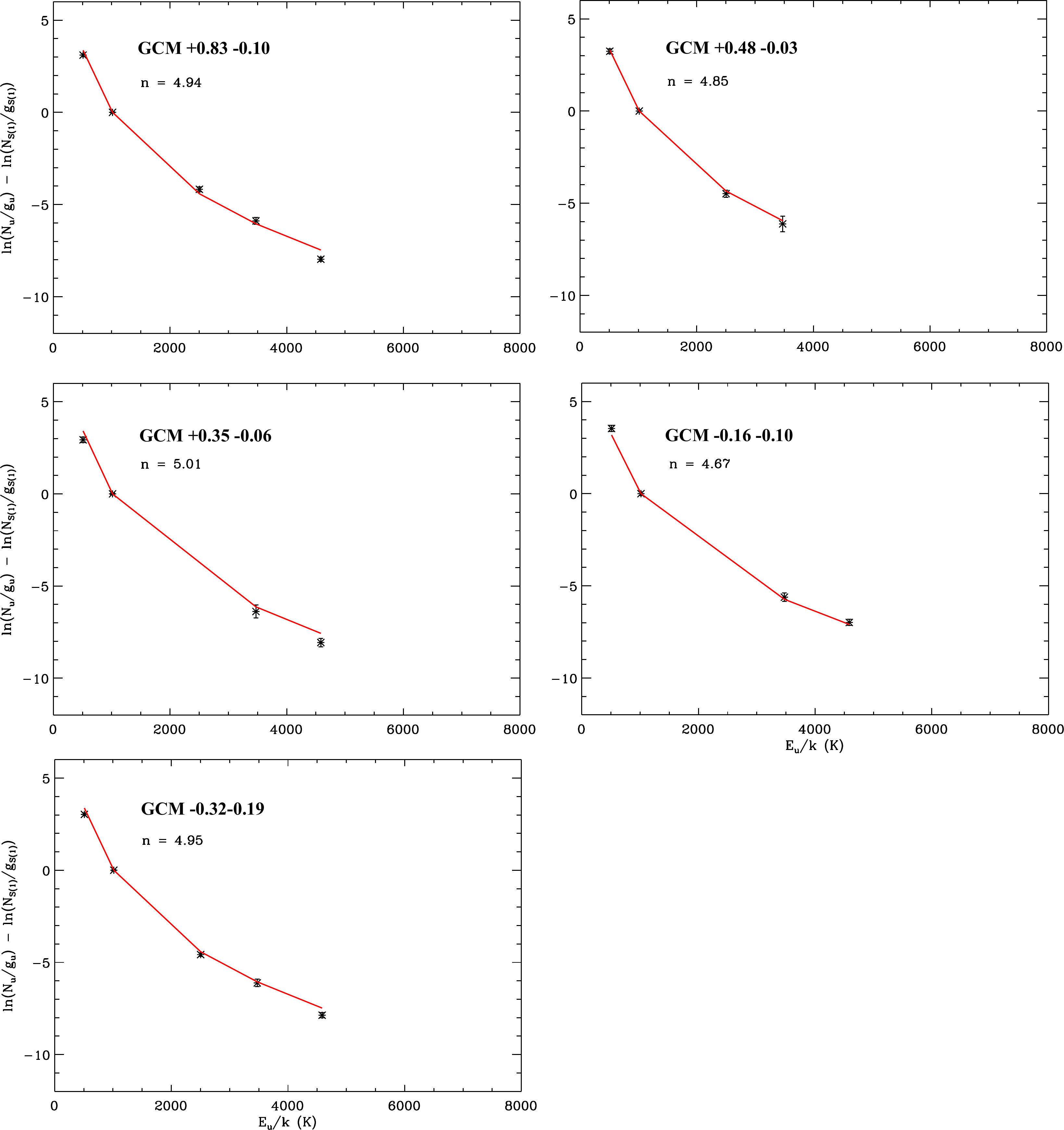}
\caption{Boltzmann plots of the observed \htwo\, lines (0-0 S) toward five Galactic center clouds from the observations of \cite{RF01}. The plotted column densities have been calculated from line fluxes that have been corrected for extinction and scaled to the observed aperture size for each line. Fits to the 0-0 S column densities for a power-law distribution of temperatures are shown.}
\vspace{-0.7cm}	
\label{fig10}
\end{figure}
\clearpage

\renewcommand\thefigure{11}
\begin{figure}[tbh]
\includegraphics[scale=0.375]{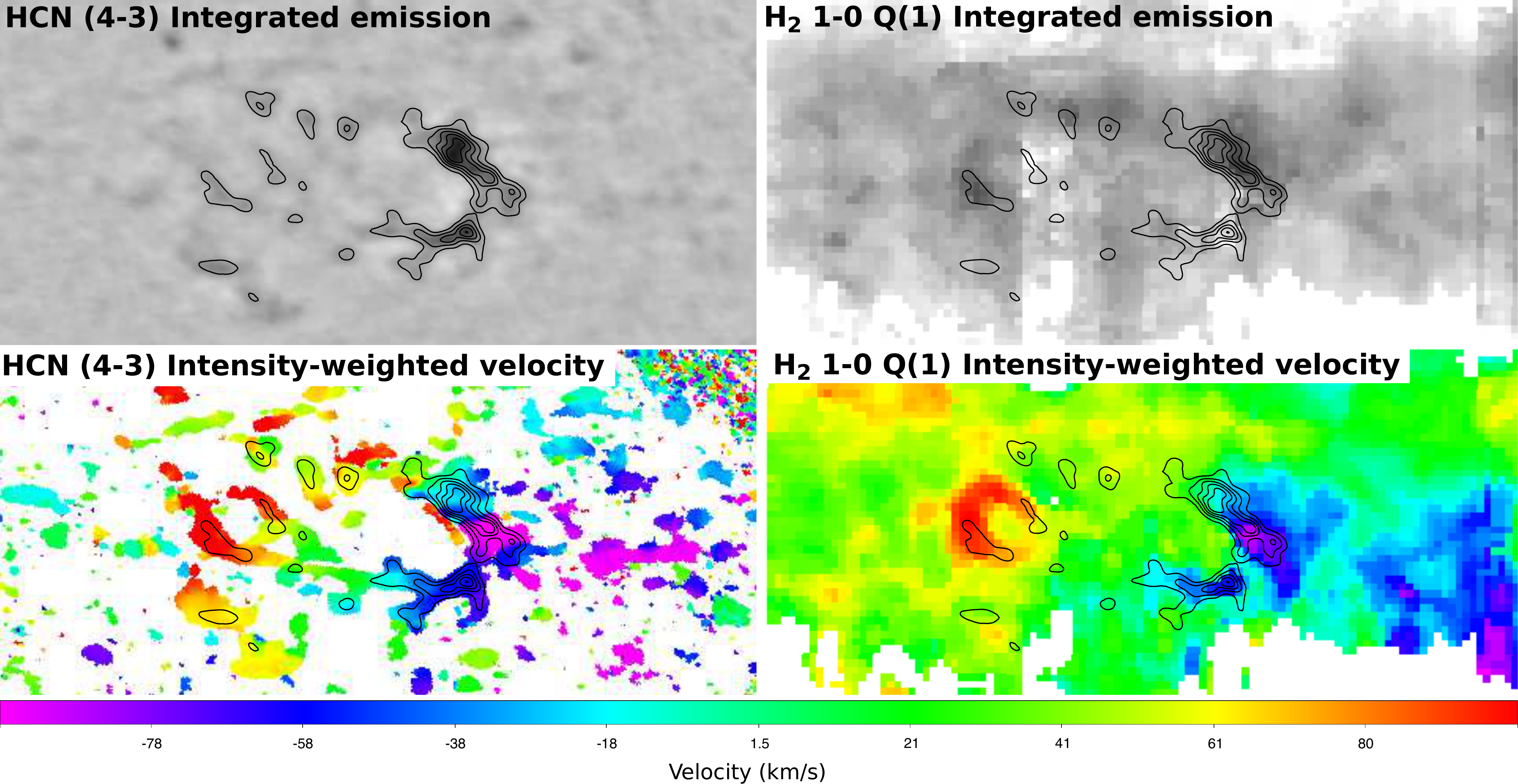}
\caption{ A comparison of emission from HCN $J=4-3$ \citep[left,][]{MCHH09} and \htwo\, 1-0 Q(1) transitions \citep[right,][]{Feldmeier14} toward the CND, oriented to be in Galactic coordinates. Top row: Integrated emission (moment 0) map. Bottom row: Intensity-weighted velocity (moment 1). Contours are the HCN $J=4-3$ integrated emission. }
\vspace{-0.7cm}	
\label{fig11}
\end{figure}
\clearpage

\begin{table}[ht]
\caption{Fitted Line Parameters for Southwest region} 
\centering
\begin{tabular}{ccccccccrc}
\\[0.5ex]
\hline\hline
& {\bf $\lambda$} & {\bf E$_{upper}$} & {\bf g$_{J}$} & {\bf Flux}  & {\bf Central}\footnotemark[1]  & {\bf Velocity}\footnotemark[2] & {\bf Integrated Flux}\footnotemark[3] & {\bf N(H2)}\footnotemark[3] & A$_\lambda$\\
& & & & & {\bf velocity}  & {\bf FWHM}  & & & \\
& ($\mu$m) & (K) & & (Jy) & (km s$^{-1}$)   & (km s$^{-1}$)  & (Jy km s$^{-1}$) & (cm $^{-2}$) & \\
& & & & & & & & & \\
\hline
\hline
 0-0 S(0) & 28.221 & 510 & 5 & $<$ 43.8 & & & $<$ 4659 & $<$ 1.5$\times 10^{21}$ & 0.97 \\
 0-0 S(1) & 17.035 & 1015 & 21 & 12.5 $\pm$ 0.4 & -38 $\pm$ 4 & 137 $\pm$ 4 & 1825 $\pm$ 98 & 3.69 $\pm$ 0.20$\times 10^{19}$ & 1.57 \\
 0-0 S(2) & 12.279 & 1682 & 9 & 16.4 $\pm$ 0.7 & -30 $\pm$ 7 & 143 $\pm$ 7 & 2518 $\pm$ 224 & 8.77 $\pm$ 0.78$\times 10^{18}$ & 1.31 \\
 0-0 S(3) &  9.665 & 2504 & 33 &  1.3 $\pm$ 0.2 & -77 $\pm$ 20 & 149 $\pm$ 20 & 198 $\pm$ 45 & 1.94 $\pm$ 0.45$\times 10^{17}$ & 5.26 \\
 0-0 S(4) &  8.026 & 3474 & 13 & 18.8 $\pm$ 0.2 & -37 $\pm$ 2 & 152 $\pm$ 2 & 3040 $\pm$ 69 & 1.11 $\pm$ 0.03$\times 10^{18}$ & 1.00 \\
 0-0 S(5) &  6.909 & 4586 & 45 & 34.2 $\pm$ 0.3 & -17 $\pm$ 1 & 178 $\pm$ 1 & 6491 $\pm$ 110 & 1.06 $\pm$ 0.02$\times 10^{18}$ & 0.88 \\
 0-0 S(6) &  6.109 & 5829 & 17 &  6.6 $\pm$ 0.4 & -38 $\pm$ 11 & 174 $\pm$ 11 & 1220 $\pm$ 120 & 1.03 $\pm$ 0.10$\times 10^{17}$ & 1.07 \\
 0-0 S(7) &  5.511 & 7197 & 57 & 17.1 $\pm$ 0.4 & -76 $\pm$ 5 & 220 $\pm$ 5 & 4020 $\pm$ 162 & 1.93 $\pm$ 0.08$\times 10^{17}$ & 0.93 \\
 0-0 S(8) &  5.053 & 8677 & 21 &  5.0 $\pm$ 0.4 & -23 $\pm$ 12 & 121 $\pm$ 12 & 647 $\pm$ 105 & 1.92 $\pm$ 0.31$\times 10^{16}$ & 0.98 \\
 0-0 S(9) &  4.695 & 10263 & 69 &  5.6 $\pm$ 0.4 & -28 $\pm$ 11 & 134 $\pm$ 11 & 792 $\pm$ 107 & 1.55 $\pm$ 0.21$\times 10^{16}$ & 1.26 \\
 0-0 S(10) &  4.410 & 11940 & 25 & $<$ 4.2 & & & $<$ 449 & $<$ 6.1$\times 10^{15}$ & 1.05 \\
 0-0 S(11) &  4.181 & 13703 & 81 &  2.1 $\pm$ 0.4 & 0 $\pm$ 32 & 133 $\pm$ 32 & 293 $\pm$ 111 & 2.92 $\pm$ 1.12$\times 10^{15}$ & 1.11 \\
 0-0 S(12) &  3.995 & 15549 & 29 &  0.5 $\pm$ 0.1 & 98 $\pm$ 40 & 161 $\pm$ 40 & 85 $\pm$ 34 & 6.45 $\pm$ 2.62$\times 10^{14}$ & 1.13 \\
 0-0 S(13) &  3.846 & 17458 & 93 &  0.8 $\pm$ 0.1 & -76 $\pm$ 26 & 155 $\pm$ 26 & 126 $\pm$ 34 & 7.52 $\pm$ 2.03$\times 10^{14}$ & 1.19 \\
 0-0 S(14) &  3.724 & 19402 & 33 & $<$ 1.0 & & & $<$ 110 & $<$ 5.3$\times 10^{14}$ & 1.27 \\
 0-0 S(15) &  3.625 & 21400 & 105 & $<$ 1.1 & & & $<$ 112 & $<$ 4.5$\times 10^{14}$ & 1.36 \\
\hline
 1-0 Q(1) &  2.407 & 6149 & 9 &  0.6 $\pm$ 0.0 & -31 $\pm$ 7 & 109 $\pm$ 7 & 73 $\pm$ 8 & 1.64 $\pm$ 1.64$\times 10^{15}$ & 2.21 \\
 1-0 Q(2) &  2.413 & 6471 & 5 & $<$ 0.3 & & & $<$ 50 & $<$ 1.6$\times 10^{15}$ & 2.20 \\
 1-0 Q(3) &  2.424 & 6956 & 21 &  0.8 $\pm$ 0.0 & -51 $\pm$ 5 & 108 $\pm$ 5 & 90 $\pm$ 7 & 3.14 $\pm$ 3.14$\times 10^{15}$ & 2.18 \\
 1-0 Q(4) &  2.438 & 7584 & 9 & $<$ 0.3 & & & $<$ 47 & $<$ 1.7$\times 10^{15}$ & 2.16 \\
 1-0 Q(5) &  2.455 & 8365 & 33 &  0.5 $\pm$ 0.0 & -49 $\pm$ 8 & 122 $\pm$ 8 & 62 $\pm$ 6 & 2.36 $\pm$ 2.36$\times 10^{15}$ & 2.13 \\
 1-0 Q(6) &  2.476 & 9286 & 13 & $<$ 0.2 & & & $<$ 32 & $<$ 1.3$\times 10^{15}$ & 2.09 \\
 1-0 Q(7) &  2.500 & 10341 & 45 & $<$ 0.2 & & & $<$ 32 & $<$ 1.3$\times 10^{15}$ & 2.05 \\
\hline
 1-0 O(2) &  2.627 & 5987 & 1 & $<$   0 & & & $<$ 74 & $<$ 8.4$\times 10^{14}$ & 1.85 \\
 1-0 O(3) &  2.803 & 6149 & 9 &  1.0 $\pm$ 0.0 & -24 $\pm$ 6 & 175 $\pm$ 6 & 182 $\pm$ 10 & 4.14 $\pm$ 0.24$\times 10^{15}$ & 1.89 \\
 1-0 O(4) &  3.004 & 6471 & 5 &  0.2 $\pm$ 0.0 & -60 $\pm$ 14 & 129 $\pm$ 14 & 25 $\pm$ 4 & 8.61 $\pm$ 1.57$\times 10^{14}$ & 2.28 \\
 1-0 O(5) &  3.235 & 6956 & 21 &  0.7 $\pm$ 0.0 & -27 $\pm$ 3 & 123 $\pm$ 3 & 94 $\pm$ 4 & 4.33 $\pm$ 0.21$\times 10^{15}$ & 1.85 \\
 1-0 O(6) &  3.501 & 7584 & 9 & $<$   1 & & & $<$ 89 & $<$ 5.8$\times 10^{15}$ & 1.58 \\
 1-0 O(7) &  3.808 & 8365 & 33 &  0.6 $\pm$ 0.1 & -62 $\pm$ 29 & 126 $\pm$ 29 & 82 $\pm$ 30 & 7.50 $\pm$ 2.73$\times 10^{15}$ & 1.20 \\
 1-0 O(8) &  4.162 & 9286 & 13 & $<$   4 & & & $<$ 630 & $<$ 8.2$\times 10^{16}$ & 1.08 \\
\hline
\end{tabular}
\label{S00}
\footnotetext[1]{Uncertainty in the wavelength calibration of ISO observations ranges from 25-60 \kms \citep{Valentijn96}}
\footnotetext[2]{The spectral resolution of SWS observations ranges from 100-300 \kms \citep{Valentijn96}}
\footnotetext[3]{Values are not corrected for extinction.}
\end{table}
\clearpage

\begin{table}[ht]
\caption{Fitted Line Parameters for Northeast region} 
\centering
\begin{tabular}{ccccccccrc}
\\[0.5ex]
\hline\hline
& {\bf $\lambda$} & {\bf E$_{upper}$} & {\bf g$_{J}$} & {\bf Flux}  & {\bf Central}\footnotemark[1]   & {\bf Velocity}\footnotemark[2]   & {\bf Integrated Flux}\footnotemark[3] & {\bf N(H2)}\footnotemark[3] & A$_\lambda$\\
& & & & & {\bf velocity}  & {\bf FWHM}  & & & \\
& ($\mu$m) & (K) & & (Jy) & (km s$^{-1}$)   & (km s$^{-1}$)  & (Jy km s$^{-1}$) & (cm $^{-2}$) & \\
& & & & & & & & & \\
\hline
\hline
 0-0 S(0) & 28.221 & 510 & 5 & $<$ 36.5 & & & $<$ 3884 & $<$ 1.3$\times 10^{21}$ & 1.16 \\
 0-0 S(1) & 17.035 & 1015 & 21 & 18.8 $\pm$ 0.4 & 7 $\pm$ 3 & 131 $\pm$ 3 & 2620 $\pm$ 111 & 5.29 $\pm$ 0.22$\times 10^{19}$ & 1.88 \\
 0-0 S(2) & 12.279 & 1682 & 9 & 18.9 $\pm$ 0.9 & 25 $\pm$ 5 & 101 $\pm$ 5 & 2034 $\pm$ 176 & 7.09 $\pm$ 0.62$\times 10^{18}$ & 1.56 \\
 0-0 S(3) &  9.665 & 2504 & 33 &  1.2 $\pm$ 0.2 & 13 $\pm$ 18 & 82 $\pm$ 18 & 106 $\pm$ 35 & 1.04 $\pm$ 0.35$\times 10^{17}$ & 6.28 \\
 0-0 S(4) &  8.026 & 3474 & 13 & 19.3 $\pm$ 0.2 & 46 $\pm$ 1 & 153 $\pm$ 1 & 3150 $\pm$ 49 & 1.15 $\pm$ 0.02$\times 10^{18}$ & 1.20 \\
 0-0 S(5) &  6.909 & 4586 & 45 & 40.6 $\pm$ 0.2 & 70 $\pm$ 0 & 178 $\pm$ 0 & 7720 $\pm$ 61 & 1.26 $\pm$ 0.01$\times 10^{18}$ & 1.05 \\
 0-0 S(6) &  6.109 & 5829 & 17 &  6.7 $\pm$ 0.2 & 44 $\pm$ 6 & 181 $\pm$ 6 & 1288 $\pm$ 74 & 1.09 $\pm$ 0.06$\times 10^{17}$ & 1.27 \\
 0-0 S(7) &  5.511 & 7197 & 57 & 15.8 $\pm$ 0.2 & -3 $\pm$ 4 & 229 $\pm$ 4 & 3867 $\pm$ 110 & 1.86 $\pm$ 0.05$\times 10^{17}$ & 1.11 \\
 0-0 S(8) &  5.053 & 8677 & 21 &  5.1 $\pm$ 0.2 & 86 $\pm$ 6 & 134 $\pm$ 6 & 731 $\pm$ 54 & 2.17 $\pm$ 0.16$\times 10^{16}$ & 1.16 \\
 0-0 S(9) &  4.695 & 10263 & 69 &  7.2 $\pm$ 0.2 & 69 $\pm$ 4 & 126 $\pm$ 4 & 963 $\pm$ 54 & 1.89 $\pm$ 0.11$\times 10^{16}$ & 1.51 \\
 0-0 S(10) &  4.410 & 11940 & 25 &  1.1 $\pm$ 0.2 & 32 $\pm$ 24 & 112 $\pm$ 24 & 134 $\pm$ 44 & 1.84 $\pm$ 0.61$\times 10^{15}$ & 1.25 \\
 0-0 S(11) &  4.181 & 13703 & 81 &  2.6 $\pm$ 0.2 & 66 $\pm$ 12 & 136 $\pm$ 12 & 377 $\pm$ 56 & 3.76 $\pm$ 0.56$\times 10^{15}$ & 1.33 \\
 0-0 S(12) &  3.995 & 15549 & 29 &  0.5 $\pm$ 0.1 & 167 $\pm$ 66 & 191 $\pm$ 66 & 94 $\pm$ 54 & 7.18 $\pm$ 4.13$\times 10^{14}$ & 1.35 \\
 0-0 S(13) &  3.846 & 17458 & 93 &  0.9 $\pm$ 0.1 & 8 $\pm$ 34 & 175 $\pm$ 34 & 161 $\pm$ 51 & 9.56 $\pm$ 3.08$\times 10^{14}$ & 1.42 \\
 0-0 S(14) &  3.724 & 19402 & 33 & $<$ 1.5 & & & $<$ 155 & $<$ 7.5$\times 10^{14}$ & 1.52 \\
 0-0 S(15) &  3.625 & 21400 & 105 & $<$ 1.5 & & & $<$ 156 & $<$ 6.2$\times 10^{14}$ & 1.62 \\
\hline
1-0 Q(1) &  2.407 & 6149 & 9 &  1.0 $\pm$ 0.0 & 87 $\pm$ 4 & 94 $\pm$ 4 & 95 $\pm$ 6 & 2.14 $\pm$ 2.14$\times 10^{15}$ & 2.63 \\
 1-0 Q(2) &  2.413 & 6471 & 5 & $<$ 0.3 & & & $<$ 43 & $<$ 1.4$\times 10^{15}$ & 2.62 \\
 1-0 Q(3) &  2.424 & 6956 & 21 &  0.8 $\pm$ 0.0 & 77 $\pm$ 4 & 112 $\pm$ 4 & 94 $\pm$ 6 & 3.27 $\pm$ 3.27$\times 10^{15}$ & 2.60 \\
 1-0 Q(4) &  2.438 & 7584 & 9 &  0.3 $\pm$ 0.0 & 32 $\pm$ 9 & 83 $\pm$ 9 & 24 $\pm$ 4 & 8.75 $\pm$ 8.75$\times 10^{14}$ & 2.57 \\
 1-0 Q(5) &  2.455 & 8365 & 33 &  0.8 $\pm$ 0.0 & 66 $\pm$ 3 & 136 $\pm$ 3 & 113 $\pm$ 5 & 4.28 $\pm$ 4.28$\times 10^{15}$ & 2.54 \\
 1-0 Q(6) &  2.476 & 9286 & 13 &  0.3 $\pm$ 0.0 & 86 $\pm$ 7 & 74 $\pm$ 7 & 22 $\pm$ 3 & 8.80 $\pm$ 8.80$\times 10^{14}$ & 2.50 \\
 1-0 Q(7) &  2.500 & 10341 & 45 &  0.5 $\pm$ 0.0 & 100 $\pm$ 6 & 122 $\pm$ 6 & 67 $\pm$ 5 & 2.75 $\pm$ 2.75$\times 10^{15}$ & 2.45 \\
\hline
 1-0 O(2) &  2.627 & 5987 & 1 & $<$   0 & & & $<$ 47 & $<$ 5.4$\times 10^{14}$ & 2.20 \\
 1-0 O(3) &  2.803 & 6149 & 9 &  0.9 $\pm$ 0.0 & 71 $\pm$ 3 & 123 $\pm$ 3 & 120 $\pm$ 5 & 2.75 $\pm$ 0.13$\times 10^{15}$ & 2.26 \\
 1-0 O(4) &  3.004 & 6471 & 5 &  0.2 $\pm$ 0.0 & 76 $\pm$ 11 & 99 $\pm$ 11 & 23 $\pm$ 3 & 7.70 $\pm$ 1.32$\times 10^{14}$ & 2.72 \\
 1-0 O(5) &  3.235 & 6956 & 21 &  0.7 $\pm$ 0.0 & 68 $\pm$ 3 & 111 $\pm$ 3 & 86 $\pm$ 4 & 4.00 $\pm$ 0.19$\times 10^{15}$ & 2.21 \\
 1-0 O(6) &  3.501 & 7584 & 9 & $<$   0 & & & $<$ 48 & $<$ 3.1$\times 10^{15}$ & 1.89 \\
 1-0 O(7) &  3.808 & 8365 & 33 &  0.7 $\pm$ 0.2 & 18 $\pm$ 42 & 151 $\pm$ 42 & 105 $\pm$ 47 & 9.53 $\pm$ 4.29$\times 10^{15}$ & 1.44 \\
 1-0 O(8) &  4.162 & 9286 & 13 & $<$   2 & & & $<$ 320 & $<$ 4.2$\times 10^{16}$ & 1.29 \\
\hline
\end{tabular}
\label{Q10}
\footnotetext[1]{Uncertainty in the wavelength calibration of ISO observations ranges from 25-60 \kms \citep{Valentijn96}}
\footnotetext[2]{The spectral resolution of SWS observations ranges from 100-300 \kms \citep{Valentijn96}}
\footnotetext[3]{Values are not corrected for extinction.}
\end{table}
\clearpage

\begin{table}[ht]
\caption{Fitted Line Parameters for Sgr A*} 
\centering
\begin{tabular}{ccccccccrc}
\\[0.5ex]
\hline\hline
& {\bf $\lambda$} & {\bf E$_{upper}$} & {\bf g$_{J}$} & {\bf Flux}  & {\bf Central}\footnotemark[1]   & {\bf Velocity}\footnotemark[2]   & {\bf Integrated Flux}\footnotemark[3] & {\bf N(H2)}\footnotemark[3] & A$_\lambda$\\
& & & & & {\bf velocity}  & {\bf FWHM}  & & & \\
& ($\mu$m) & (K) & & (Jy) & (km s$^{-1}$)   & (km s$^{-1}$)  & (Jy km s$^{-1}$) & (cm $^{-2}$) & \\
& & & & & & & & & \\
\hline
\hline
 0-0 S(0) & 28.221 & 510 & 5 & $<$ 124.9 & & & $<$ 13290 & $<$ 4.3$\times 10^{21}$ & 0.99 \\
 0-0 S(1) & 17.035 & 1015 & 21 & $<$ 11.9 & & & $<$ 1265 & $<$ 2.6$\times 10^{19}$ & 1.61 \\
 0-0 S(2) & 12.279 & 1682 & 9 & $<$ 68.3 & & & $<$ 7273 & $<$ 2.5$\times 10^{19}$ & 1.33 \\
 0-0 S(3) &  9.665 & 2504 & 33 & $<$ 6.6 & & & $<$ 698 & $<$ 6.8$\times 10^{17}$ & 5.37 \\
 0-0 S(4) &  8.026 & 3474 & 13 &  9.6 $\pm$ 1.3 & 2 $\pm$ 29 & 186 $\pm$ 29 & 1910 $\pm$ 497 & 6.96 $\pm$ 1.81$\times 10^{17}$ & 1.02 \\
 0-0 S(5) &  6.909 & 4586 & 45 & 30.6 $\pm$ 1.8 & 36 $\pm$ 12 & 189 $\pm$ 12 & 6175 $\pm$ 662 & 1.01 $\pm$ 0.11$\times 10^{18}$ & 0.90 \\
 0-0 S(6) &  6.109 & 5829 & 17 &  7.1 $\pm$ 1.7 & 0 $\pm$ 40 & 147 $\pm$ 40 & 1110 $\pm$ 460 & 9.37 $\pm$ 3.89$\times 10^{16}$ & 1.09 \\
 0-0 S(7) &  5.511 & 7197 & 57 & 12.0 $\pm$ 1.2 & 53 $\pm$ 29 & 264 $\pm$ 29 & 3385 $\pm$ 629 & 1.63 $\pm$ 0.30$\times 10^{17}$ & 0.95 \\
 0-0 S(8) &  5.053 & 8677 & 21 &  4.7 $\pm$ 1.0 & 4 $\pm$ 25 & 105 $\pm$ 25 & 531 $\pm$ 198 & 1.58 $\pm$ 0.59$\times 10^{16}$ & 1.00 \\
 0-0 S(9) &  4.695 & 10263 & 69 & $<$ 8.2 & & & $<$ 871 & $<$ 1.7$\times 10^{16}$ & 1.29 \\
 0-0 S(10) &  4.410 & 11940 & 25 & $<$ 8.0 & & & $<$ 847 & $<$ 1.2$\times 10^{16}$ & 1.07 \\
 0-0 S(11) &  4.181 & 13703 & 81 & $<$ 9.6 & & & $<$ 1021 & $<$ 1.0$\times 10^{16}$ & 1.13 \\
 0-0 S(12) &  3.995 & 15549 & 29 & $<$ 10.9 & & & $<$ 1157 & $<$ 8.8$\times 10^{15}$ & 1.15 \\
 0-0 S(13) &  3.846 & 17458 & 93 & $<$ 10.9 & & & $<$ 1157 & $<$ 6.9$\times 10^{15}$ & 1.21 \\
 0-0 S(14) &  3.724 & 19402 & 33 & $<$ 10.9 & & & $<$ 1156 & $<$ 5.6$\times 10^{15}$ & 1.30 \\
 0-0 S(15) &  3.625 & 21400 & 105 & $<$ 10.9 & & & $<$ 1156 & $<$ 4.6$\times 10^{15}$ & 1.39 \\
\hline
1-0 Q(1) &  2.407 & 6149 & 9 & $<$ 0.7 & & & $<$ 108 & $<$ 2.4$\times 10^{15}$ & 2.25 \\
 1-0 Q(2) &  2.413 & 6471 & 5 & $<$ 0.6 & & & $<$ 94 & $<$ 3.0$\times 10^{15}$ & 2.24 \\
 1-0 Q(3) &  2.424 & 6956 & 21 & $<$ 0.5 & & & $<$ 79 & $<$ 2.8$\times 10^{15}$ & 2.22 \\
 1-0 Q(4) &  2.438 & 7584 & 9 & $<$ 0.4 & & & $<$ 68 & $<$ 2.5$\times 10^{15}$ & 2.20 \\
 1-0 Q(5) &  2.455 & 8365 & 33 & $<$ 0.4 & & & $<$ 60 & $<$ 2.3$\times 10^{15}$ & 2.17 \\
 1-0 Q(6) &  2.476 & 9286 & 13 & $<$ 0.3 & & & $<$ 54 & $<$ 2.1$\times 10^{15}$ & 2.14 \\
 1-0 Q(7) &  2.500 & 10341 & 45 & $<$ 0.4 & & & $<$ 58 & $<$ 2.4$\times 10^{15}$ & 2.10 \\
\hline
 1-0 O(2) &  2.627 & 5987 & 1 & $<$   0 & & & $<$ 60 & $<$ 6.9$\times 10^{14}$ & 1.88 \\
 1-0 O(3) &  2.803 & 6149 & 9 & $<$   0 & & & $<$ 48 & $<$ 1.1$\times 10^{15}$ & 1.93 \\
 1-0 O(4) &  3.004 & 6471 & 5 & $<$   0 & & & $<$ 37 & $<$ 1.2$\times 10^{15}$ & 2.33 \\
 1-0 O(5) &  3.235 & 6956 & 21 &  0.3 $\pm$ 0.1 & 34 $\pm$ 22 & 82 $\pm$ 22 & 24 $\pm$ 10 & 1.13 $\pm$ 0.47$\times 10^{15}$ & 1.89 \\
 1-0 O(6) &  3.501 & 7584 & 9 &  0.3 $\pm$ 0.1 & 66 $\pm$ 36 & 114 $\pm$ 36 & 39 $\pm$ 25 & 2.54 $\pm$ 1.62$\times 10^{15}$ & 1.61 \\
 1-0 O(7) &  3.808 & 8365 & 33 & $<$  11 & & & $<$ 1735 & $<$ 1.6$\times 10^{17}$ & 1.23 \\
 1-0 O(8) &  4.162 & 9286 & 13 & $<$  10 & & & $<$ 1563 & $<$ 2.0$\times 10^{17}$ & 1.10 \\
\hline
\end{tabular}
\label{O10}
\footnotetext[1]{Uncertainty in the wavelength calibration of ISO observations ranges from 25-60 \kms \citep{Valentijn96}}
\footnotetext[2]{The spectral resolution of SWS observations ranges from 100-300 \kms \citep{Valentijn96}}
\footnotetext[3]{Values are not corrected for extinction.}
\end{table}
\clearpage

\begin{table}[ht]
\caption{Fraction of Hot H$_2$} 
\centering
\begin{tabular}{llll}
\\[0.5ex]
\hline\hline
& & & \\
{\bf T$_{ex,i}$}  & {\bf C$_{i}$\footnotemark[1]}  & {\bf N$_{\mathrm{H2},i}$}  & {\bf fraction of} \\
(K) & (cm$^{-2}$)   & (cm$^{-2}$)  & {\bf hot H$_2$} \\
\hline
\multicolumn{4}{c}{Southwest region} \\
\hline
580   &  4.20$\times10^{19}$  & 6.02$\times10^{20}$  &  93.68\%\\
1350 &  1.19$\times10^{18}$  & 4.03$\times10^{19}$ &    6.27\%\\
3630 &  2.92$\times10^{15}$  &  3.24$\times10^{17}$ &   0.05\%\\
\hline
\multicolumn{4}{c}{Northeast region} \\
\hline
520   &  8.82$\times10^{19}$  & 1.13$\times10^{21}$   &  94.14\%\\
1260 &  2.21$\times10^{18}$  & 6.93$\times10^{19}$ &    5.76\%\\
2840 &  1.56$\times10^{16}$  & 1.24$\times10^{18}$ &    0.10\%\\
\hline
\end{tabular}
\label{fraction}
\footnotetext[1]{See Equation \ref{eq1b}.}
\end{table}
\clearpage

\begin{table}[ht]
\caption{CND Masses} 
\centering
\begin{tabular}{lcll}
\\[0.5ex]
\hline\hline
& & & \\
{\bf Tracer}  & {\bf Aperture} & \multicolumn{2}{c}{\bf Total Molecular Mass\footnotemark[1]} \\
& ($''\times''$) & \multicolumn{2}{c}{(M$_{\odot}$)} \\
& & {Southwest region} & {Northeast region}\\
\hline
\htwo\, (Discrete temperature fit, T$>$ 500 K)  &  14$\times$20  & 6.1 & 11.4\\
\htwo\, (Power law extrapolation, T$>$ 100 K)  &  14$\times$20  & 41 & 111 \\
\htwo\, (Power law extrapolation, T$>$ 50 K)  &  14$\times$20  & 146  & 518 \\
Dust continuum (Battersby et al, in. prep) &  14$\times$20  & 456  & 370  \\
CO \citep{RT12} &  22 &  557\footnotemark[2] & 413\footnotemark[2] \\
\hline
\end{tabular}
\label{masses}
\footnotetext[1]{Masses given are the total molecular mass, assuming a mean molecular weight of 2.8, consistent with the Galactic center metallicity.}
\footnotetext[2]{Scaled to an area equivalent to a $14''\times20''$ aperture, assuming a uniform flux distribution}
\end{table}
\clearpage


\end{document}